\documentclass[]{biometrika}

\usepackage{amsmath}

\usepackage{times}
\usepackage{bm}
\usepackage{xfrac}
\usepackage{natbib}
\usepackage{amssymb}
\usepackage{amsmath}
\usepackage[plain,noend]{algorithm2e}
\usepackage[export]{adjustbox}
\usepackage{chngcntr}

\usepackage{graphicx,epsfig,amscd,array}
\usepackage{subfigure}
\newcommand{\innerprod}[2]{\langle#1,#2\rangle}
\newcommand{\norm}[1]{\lVert#1\rVert}
\newtheorem{conjecture}{Conjecture}

\makeatletter
\renewcommand{\algocf@captiontext}[2]{#1\algocf@typo. \AlCapFnt{}#2} 
\def\@algocf@capt@plain{top}
\renewcommand{\algocf@makecaption}[2]{%
  \addtolength{\hsize}{\algomargin}%
  \sbox\@tempboxa{\algocf@captiontext{#1}{#2}}%
  \ifdim\wd\@tempboxa >\hsize
    \hskip .5\algomargin%
    \parbox[t]{\hsize}{\algocf@captiontext{#1}{#2}}
  \else%
    \global\@minipagefalse%
    \hbox to\hsize{\box\@tempboxa}
  \fi%
  \addtolength{\hsize}{-\algomargin}%
}
\makeatother

\begin{document}

\jname{Biometrika}
\jyear{20??}
\jvol{??}
\jnum{?}
\accessdate{Advance Access publication on ?? ???? 20??}
\copyrightinfo{\Copyright\ 20?? Biometrika Trust\goodbreak {\em Printed in Great Britain}}

\received{? 20??}
\revised{? 20??}

\markboth{D. Simpson, J. Illian, F. Lindgren, S. H. S\o rbye \and H. Rue}{Efficient inference for log-Gaussian Cox processes}

\title{Going off grid:  Computationally efficient inference for log-Gaussian Cox processes}


\author{D. SIMPSON}
\affil{Department of Mathematical Sciences, University of Bath, Bath BA2 7AY, U.K. \email{d.simpson@bath.ac.uk}}

\author{J. B. ILLIAN}
\affil{Centre for Research into Ecological and Environmental Modelling, University of St Andrews, St Andrews, Fife KY16 9LZ,  U.K. \\ Department of Mathematical Sciences, Norwegian University of Science and Technology, N-7491 Trondheim, Norway. \email{jbi@st-andrews.ac.uk}}

\author{F. LINDGREN}
\affil{Department of Mathematical Sciences, University of Bath, Bath BA2 7AY, U.K. \email{f.lindgren@bath.ac.uk}}

\author{S. H. S\O RBYE}
\affil{Department of Mathematics and Statistics, UiT The Arctic University of Norway, N-9037 Troms{\o}, Norway \email{sigrunn.sorbye@uit.no}}

\author{\and H. RUE}
\affil{Department of Mathematical Sciences, Norwegian University of Science and Technology, N-7491 Trondheim, Norway  \email{hrue@math.ntnu.no}}

\maketitle

\begin{abstract}
This paper introduces a new method for performing computational inference on log-Gaussian Cox processes. The likelihood is approximated directly by making novel use of a continuously specified Gaussian random field. We show that for sufficiently smooth Gaussian random field prior distributions, the approximation can converge with arbitrarily high order, while an approximation based on a counting process on a partition of the domain only achieves first-order convergence.  The given results improve on the general theory of convergence of the stochastic partial differential equation models, introduced by \citet{Lindgren2011}.  The new method is demonstrated on a standard point pattern data set and two interesting extensions to the classical log-Gaussian Cox process framework are discussed. The first extension considers variable sampling effort throughout the observation window and implements the method of \citet{Chakraborty2011}.  The second extension constructs a log-Gaussian Cox process on the world's oceans. The analysis is performed using integrated nested Laplace approximation  
for fast approximate inference.  
\end{abstract}


\section{Introduction}

Data consisting of sets of locations at which
some objects  are present are common  in biology, ecology and economics.  The appropriate statistical models for this
type of data are spatial point process models, which have
been extensively studied by statisticians and probabilists 
\citep{book87,illianBook} but are less commonly used by the scientists producing the data sets.  Point process models are often  hard to fit, so scientists often resort to using inappropriate methods. \citep{Chakraborty2011}  discuss this in the context of presence-only datasets, and outline various ad hoc approaches used by ecologistsThere is an interesting discussion of this in the context of presence only data sets \citep{Chakraborty2011}, which outlines a number of ad hoc approaches taken by the ecological community.

Many real data sets do not have the simple structure usually considered in the classical statistical literature, i.e., that of a simple point pattern that has been observed everywhere within a simple, often rectangular, plot.  For instance, in real data sets the observation process is often not straightforward due to practical limitations, or the observation window  is  complex. This includes 
data sets mapping the locations of bird species, for which very little data have been collected in the Himalayas for obvious reasons.  Therefore, on
top of sampling issues such as incompletely observed point
patterns, positional errors, etc., this data set has a large hole where it is believed that birds reside, but it is impractical to look for them.  Very different, but similarly complex, data deal with freak waves in the oceans.  Even if we ignore  temporal aspects, or the uncertainty in  the observed locations, this data set remains complicated,  as the observation window covers most of a sphere and has a very complicated boundary.   Motivated by data sets of this nature, this paper
proposes an easy to use, computationally efficient method for
performing inference on spatial point process models that is
sufficiently flexible to handle these and other data structures.

In this paper we focus on log-Gaussian Cox processes, 
a class of flexible models that is particularly useful in the context
of modelling aggregation relative to some underlying  unobserved environmental field
\citep{art245,illianalb:11}. However, standard methods for fitting Cox
processes are computationally expensive and the Markov chain Monte Carlo methods that are
commonly used are difficult to tune for this problem.
Recently, \citet{illianalb:11} developed a fast, flexible framework for
fitting log-Gaussian Cox processes  using integrated nested Laplace
approximation \citep{art451}.  They construct a
Poisson approximation to the true log-Gaussian Cox process likelihood to perform inference on a regular lattice over the observation window, counting the
number of points in each cell.  If the lattice is fine enough and the latent Gaussian field is appropriately
discretised, this approximation is quite good
\citep{waagepetersen2004}, but it can be computationally wasteful,
especially when the process intensity is high or the
observation window is large or oddly shaped.  New results on the strong convergence of the lattice approximation, provided in the Appendix, show that the rate of convergence on a $p\times p$ lattice is fundamentally limited to $\mathcal{O}(p^{-1})$ by the counting approximation.

In the Appendix, we provide detailed results on the convergence of the approximations proposed in this paper. In particular, we show that, for a Gaussian random field with fixed parameters, the posterior distributions generated using the  proposed method will converge strongly to the true posterior distribution.  Furthermore, it is shown that these posterior distributions  can converge with arbitrarily high order and the convergence is limited only by the smoothness of the random field. In this paper, we place particular emphasis on the combination of this method with the flexible stochastic partial differential equation models of \citet{Lindgren2011} and we significantly improve the existing convergence theory for these models.  In particular, we show that the approximate posterior distributions converge weakly and the error when computing a posterior functional is almost $\mathcal{O}(h^{-1})$.

The first of our aims is to
re-examine the standard methodology for  Bayesian inference
on log-Gaussian Cox processes  and to propose an approach that is
much more computationally efficient based on  continuously specified finite-dimensional Gaussian random fields.   The
key characteristic of our approach is that the specification of the Gaussian random field is
completely separated from the approximation of the likelihood, leading to far greater flexibility.  The second aim is to demonstrate that this approach can be handled  within the general approximation framework of \citet{art451}, by modelling the Gaussian random field through a  stochastic partial
differential equation  \citep{Lindgren2011}. This provides a unified modelling structure. An associated \texttt{R}-package 
makes our methods that accessible to scientists.  

\section{Log-Gaussian Cox processes} \label{sec:lgcp}

Consider a bounded region $\Omega \subset \mathbb{R}^2$.  A
simple point process model is the inhomogeneous
Poisson process, in which the number of points within a region $D
\subset \Omega$  is Poisson distributed with mean $\Lambda(D) = \int_D
\lambda(s)\,ds$, where $\lambda(s)$ is the intensity surface
of the point process. Given the intensity surface and a point pattern
$Y,$ the likelihood of an inhomogeneous Poisson process is
\begin{equation} \label{likelihood}
\pi({Y} \mid \lambda) = \exp\left\{|\Omega| - \int_{\Omega} \lambda(s) \,ds\right\}\prod_{s_i \in Y} \lambda(s_i).
\end{equation}
This likelihood is analytically intractable, as it requires the integral of the intensity function, which typically cannot be calculated explicitly. This integral can, however, be computed numerically using standard methods.

Treating the intensity surface as a realisation of a {random
  field} $\lambda(s)$ yields a particularly flexible class of point
processes known as {Cox} or doubly stochastic Poisson
processes \citep{book87}. These are typically used to model aggregation in point
patterns resulting from observed or unobserved environmental
variation. In this paper we consider {log-Gaussian Cox processes}, where the intensity surface is modelled as  $\log\lambda(s) = {Z}(s)$, and  ${Z}(s)$ is a Gaussian random field.  Conditional on a
realisation of ${Z}(s)$, a log-Gaussian Cox process is an
inhomogeneous Poisson process.  The likelihood for
such a process is of the form \eqref{likelihood}, where the integral is further
complicated by the stochastic nature of $\lambda(s)$, and methods for
approximating \eqref{likelihood} are the focus of the next two
sections.  Log-Gaussian Cox processes fit naturally
within the Bayesian hierarchical modelling framework and are latent Gaussian models. They may be fitted using the integrated nested Laplace approximation approach of \citet{art451},  allowing us to construct models that include covariates, marks and  non-standard observation processes while still allowing computationally efficient inference \citep{illianalb:11}.  Therefore, approximating the likelihood in \eqref{likelihood} constitutes a basic calculation for practical problems such as those discussed in Section \ref{sec:examples}.

\section{Computation on fine lattices is wasteful} \label{sec:lattice}
A common method for performing inference with log-Gaussian Cox processes  is to take the observation window $\Omega$,  construct a fine regular lattice over it, and then consider the number of points $N_{ij}$ observed in each cell $s_{ij}$ of the lattice  \citep{art245,illianalb:11}.  It is a simple consequence of the definition of a log-Gaussian Cox process that the $N_{ij}$ may be considered as independent Poisson random variables, that is  $
N_{ij} \sim \mbox{Po}(\Lambda_{ij}),
$ where $\Lambda_{ij} = \int_{s_{ij}} \lambda(s)\,ds$ is the total
intensity in each cell. It is impossible to
compute the total intensity for each cell and we therefore use the approximation $\Lambda_{ij} \approx
|s_{ij}| \exp(z_{ij})$, where $z_{ij}$ is a representative value of
${Z(s)}$ within the cell $s_{ij}$  and $|s_{ij}|$ is the area of cell
$s_{ij}$. The log-Gaussian Cox process model can then be treated  within the  generalised linear mixed model framework.
This method has been used in a number of applications
and converges to the true solution as the size of
the cells decreases to zero; see Corollary A\ref{cor:lattice} or \citet{waagepetersen2004}. 

The computational challenge is that, if $Z(s)$ is a general Gaussian random field, the multivariate Gaussian vector $z$ that contains the $z_{ij}$s will have a dense covariance matrix.  The resulting computational complexity  limits this method to quite small lattices.  If $Z(s)$ is stationary and the observation window is a rectangle, it is possible to use the block Toeplitz structure of the covariance matrix to speed up some computations \citep{art245}.  Unfortunately, the block Toeplitz structure is fragile and any inference method that constructs a second-order approximation to the posterior distribution, such as manifold Markov chain Monte Carlo simulation \citep{Girolami:2011hw}  or the integrated nested Laplace approximation, will destroy the computational savings.

A common computationally efficient approach is to model $z$ as
a conditional autoregressive  model on the fine lattice and use this to
perform fast computations \citep{book80}.  The conditional autoregressive approach has been used
extensively in applications  and may be fitted using the integrated nested Laplace approximation \citep{illianalb:11}. Both  methods rely heavily on the regularity
of the lattice, as it is quite difficult to
construct a conditional autoregressive model on an irregular lattice that is resolution-consistent \citep{book80}. 


However, these methods are unsatisfactory since the computational lattice has two
fundamentally different roles.  The first and most natural role is
to approximate the latent Gaussian random field ${Z}(s)$.
The second and rather unnatural role of the
computational lattice is to  approximate the locations of the points, even though the data have often been collected with   high   precision.
Clearly, the finer the lattice is, the less information is lost, so the quality of the likelihood approximation primarily depends on the size of the grid.  In fact, Corollary A\ref{cor:lattice}  shows that this binning process is the dominant source of error in the lattice approximation.  As a result, we are required to compute on a
much finer grid than is necessary for the approximation of the latent
Gaussian field, making lattice-based approaches inherently
 wasteful in this context.

The inflexibility inherent in lattice-based methods also implies that the approximation to the latent random field cannot be locally refined. In the problem considered in Section
\ref{sec:sampling},  a large region has not been sampled.
Generating a high resolution approximation to the latent field over this area would be computationally wasteful. It would be more efficient to reduce the resolution in these areas without affecting  that in those that have been sampled.  While this is 
impossible with lattice-based methods, the flexible method introduced here allows local changes to the resolution of the approximation.

\section{Approximating the likelihood using a finite-dimensional random field} \label{sec:approx}

Rather than defining a Gaussian random field over a fine lattice, we  propose a finite-dimensional 
continuously specified random field of the form
\begin{equation} \label{basis_expansion}
{Z}(s) = \sum_{i=1}^n z_i \phi_i(s),
\end{equation} where $z = (z_1, \ldots,z_n)^T$ is a
multivariate Gaussian random vector and  $\{\phi_i(s)\}_{i=1}^n$
is a set of linearly independent deterministic basis functions. This is similar in spirit to the Karhunen--Lo\`{e}ve decomposition of stochastic processes, which is based on eigen-decomposition of the covariance function of the process. Three other common
approximations to Gaussian random fields can also be expressed as in (\ref{basis_expansion}). Process convolution models 
\citep{art473} 
use the approximation $
Z(s)  = \int_\Omega k(s,s')dW(s') \approx \sum_{i=1}^N z_i k(s,s_i),
$ where the first integral is a white noise integral, the $z_i$ are independent Gaussian random variables, and the points $s_i$ lie on a lattice within $D$.  The second class of models uses correlated weights $z$ and selects basis functions, either based on a parent Gaussian process as for predictive processes \citep{art444}, or from other considerations, as in fixed-rank kriging \citep{art445}.  
\citet{Chakraborty2011}  investigated log-Gaussian Cox process models using predictive processes.    The third class comprises the stochastic partial differential
equation models of \citet{Lindgren2011}, which take $\phi_i(s)$ to be compactly-supported piecewise linear functions.    This choice of $\phi_i(s)$ delivers considerable computational benefits and will be further explored in Section \ref{sec:spde} and Appendix \ref{appendix:spde_converge}.  All of the examples in this paper use stochastic partial differential equation models for the latent process $Z(s)$.

With the continuous Gaussian random field model in place, we are in a position to
attack  the intractable likelihood \eqref{likelihood}.  In this section, we  outline a procedure for approximating the likelihood
that extends the standard approximation to the non-lattice, unbinned data case.  The log-likelihood $
\log\pi(y\mid Z) = |\Omega| - \int_\Omega \exp\{Z(s)\}\,ds + \sum_{i=1}^N Z(s_i)
$ consists of two terms: the stochastic integral, and the evaluation of the field at the data points. While the continuously-specified stochastic partial differential equation models allow us to compute the sum term exactly, we must approximate the  integral by a sum.  Consider a  deterministic integration rule of the general form $
\int_\Omega f(s)\, ds \approx \sum_{i=1}^p \tilde{\alpha}_i f(\tilde{s}_i),
$ for fixed, deterministic nodes $\{\tilde{s}_i\}_{i=1}^p$ and weights $\{\tilde{\alpha}_i\}_{i=1}^p$.  Using this integration rule,  we can construct the approximation
\begin{eqnarray}
 \log\{\pi(y\mid z)\} &\approx &C - \sum_{i=1}^p \tilde{\alpha}_i \exp\left\{\sum_{j=1}^n z_j \phi_j(\tilde{s}_i)\right\} + \sum_{i=1}^N \sum_{j=1}^n z_j\phi_j(s_i) \notag \\
&= &C - \tilde{\alpha}^T \exp(A_1 z) + 1^T A_2 z, \label{approx_likelihood}
\end{eqnarray}
where $C$ is a constant, $[A_1]_{ij} = \phi_j(\tilde{s}_i)$ is a matrix containing the values of the latent Gaussian model \eqref{basis_expansion} at the integration nodes $\{\tilde{s}_i\}$, and $[A_2]_{ij} = \phi_j(s_i)$ evaluates the latent Gaussian field at the observed points $\{s_i\}$.


The advantage of the approximation \eqref{approx_likelihood} is that it is of Poisson form.  In particular,  given $z$ and $\theta$, the approximate likelihood consists of $N + p$ independent Poisson random variables. To see this, we write $\log\eta = (z^T A_1^T,z^T A_2^T)^T$ and $\alpha = (\tilde{\alpha}^T,0_{N\times 1}^T)^T$.  Then, if we construct some pseudo-observations $y = (0_{p \times 1}^T , 1_{N \times 1}^T)^T$, the approximate likelihood factors as
\begin{align} \label{final_likelihood}
 \pi(y \mid z) \approx C\prod_{i=1}^{N + p}  \eta_i^{y_i} e^{-\alpha_i \eta_i},
\end{align} which is similar to  the likelihood for observing $N+p$ conditionally independent
Poisson random variables with means $\alpha_i\eta_i$ and observed values
$y_i$.

Numerical integration schemes that lead to likelihood approximations of the form \eqref{final_likelihood} were also
considered by \citet{badtur:00} for approximating  pseudolikelihoods of
Gibbs-type point processes. However, to the best of our knowledge,
these ideas have not been extended to log-Gaussian Cox processes, probably due to the
paucity of computationally efficient continuously specified Gaussian
random field models.  

In the Appendix, we show that the approximate posterior distribution converges  to the true posterior distribution generated using the correct log-Gaussian Cox process likelihood at a rate that depends on the smoothness of the field and the quality of the integration rule.  Hence, while \citet{badtur:00} suggest placing ``one [...] point, either
systematically or randomly'', for log-Gaussian Cox processes, there is a strong advantage to carefully designing the underlying integration scheme.

\section{Stochastic partial differential equations and Markov random fields} \label{sec:spde}
%

The approximation outlined in Section \ref{sec:approx} will work for any finite-dimensional random field \eqref{basis_expansion}.  This section shows how this approach fits naturally with our preferred finite-dimensional random field model.  In particular, we  review the stochastic partial differential equation construction of  \citet{Lindgren2011} and show how this naturally extends the   conditional autoregressive modelling strategy of \citet{illianalb:11}.

The basic idea of \citet{Lindgren2011} is that, given a surface, an
appropriate lower-resolution approximation to the surface can be
constructed by sampling the surface in a set of well designed points
and constructing a piecewise linear interpolant.  We will, therefore, take the basis
functions in \eqref{basis_expansion} to be a set of piecewise linear
functions defined over a triangular mesh, which gives  more
geometric flexibility than does a traditional grid-based method.

We consider Mat\'{e}rn random fields, i.e.\ zero-mean Gaussian stationary, isotropic random fields with covariance function $
c(h) = \left\{\Gamma(\nu +d/2) (4\pi)^{d/2}  2^{\nu-1} \kappa^{2\nu}\tau^2 \right\}^{-1}(\kappa h)^\nu K_\nu(\kappa h)$,  where  $h \geq 0$, $K_\nu(\cdot)$ is the modified Bessel function of the second kind, $\nu>0$ is the smoothing parameter, $\kappa >0$ is the range parameter,  $\tau$ is a scaling parameter, and the normalisation is chosen to link it with the representation \eqref{SPDE}. The subset of Mat\'{e}rn random fields for which $\nu + d/2$ is an integer, where $d$ is the dimension of the space, yields computationally efficient piecewise linear representations by  representing of the Mat\'{e}rn field $Z(s)$ as the stationary solution to the stochastic partial differential equation  
\begin{equation} \label{SPDE}
\tau(\kappa^2 - \Delta)^{\alpha/2} Z(s) = W(s),
\end{equation} where $\alpha= \nu - d/2$ is an integer, $\Delta = \sum_{i=1}^d \partial^2 /\partial s_i^2$ is the Laplacian operator, and $W(s)$ is spatial white noise.
This representation  was first constructed by \citet{art246,art455} while proving that the classical second-order conditional autoregression model converges under lattice refinement to a Mat\'{e}rn field with $\nu = 1$.

Piecewise linear approximations to deterministic partial
differential equations are commonly constructed in physics,
engineering and applied mathematics using the finite element method, which was used by 
\citet{Lindgren2011} to efficiently represent the
appropriate Mat\'{e}rn fields.  When $\alpha=2$, the final outcome of their procedure replaces the stochastic partial differential equation \eqref{SPDE} with a simple equation for the weights in the basis expansion  \eqref{basis_expansion}
\begin{equation} \label{spde_approx}
(\kappa^2 C  + G - B) z \sim N(0, C),
\end{equation}
 where $B$, $C$ and $G$ are  sparse
 matrices with entries
$$
C_{ii}= \int_\Omega \phi_i(s) \,ds,\quad 
G_{ij} = \int_\Omega \nabla \phi_i(s) \nabla \phi_j(s)\,ds, \quad
B_{ij} = \int_{\partial\Omega} \phi_i(s) \partial_n \phi_j(s)\,ds.
$$
The boundary of $\Omega$ is $\partial \Omega$, while $\partial_n\phi_j(s)$ is the normal derivative of $\phi_j(s)$ and $C$ is diagonal, see Appendix C.5 in \citet{Lindgren2011} for a discussion on the choice of $C$. \cite{Lindgren2011} also show that these models lead exactly to the classical conditional autoregressive models when computed over a regular lattice.  This model can be extended to 
non-stationary, anisotropic, multivariate and spatiotemporal  random fields \citep{Cameletti2011,fuglstad2013exploring}, and the methods described in this paper extend to these cases in a straightforward way, although the implementation of these models may be  non-trivial.

The matrix $B$ in \eqref{spde_approx} encodes information on the
process on the boundary of the observation window $\Omega$.  The
effect of physical boundaries in spatial models has received very
little attention in the literature. A notable example in the context
of Bayesian smoothing is \cite{art457}.  For the remainder of this
paper, we will set $B=0$, which corresponds to
Neumann, or no-flux, boundary conditions.  These specify that the normal derivative of the field at the boundary is zero and can be physically related to an insulating boundary from which no heat escapes.  We  discuss the interpretation of
this condition in Section \ref{sec:ocean}.

We suggest a meshing strategy that constructs a regular triangulation of the observation window, and refine it in areas where there are a large number of points. Point pattern data hold information on the relevant point process even in areas with only a few points.  Hence, in order to  avoid approximation bias introduced by the choice of mesh, the triangulation needs to cover the space in a fairly regular way.  On the other hand, we are unlikely to be able to infer the fine-scale latent structure in areas where we  have no points or there has been little sampling. A detailed discussion of mesh selection can be found in Chapter 6 of \citet{blangiardo2015spatial}.

\begin{figure}[t]
  \centering
	\includegraphics[width=0.5\textwidth]{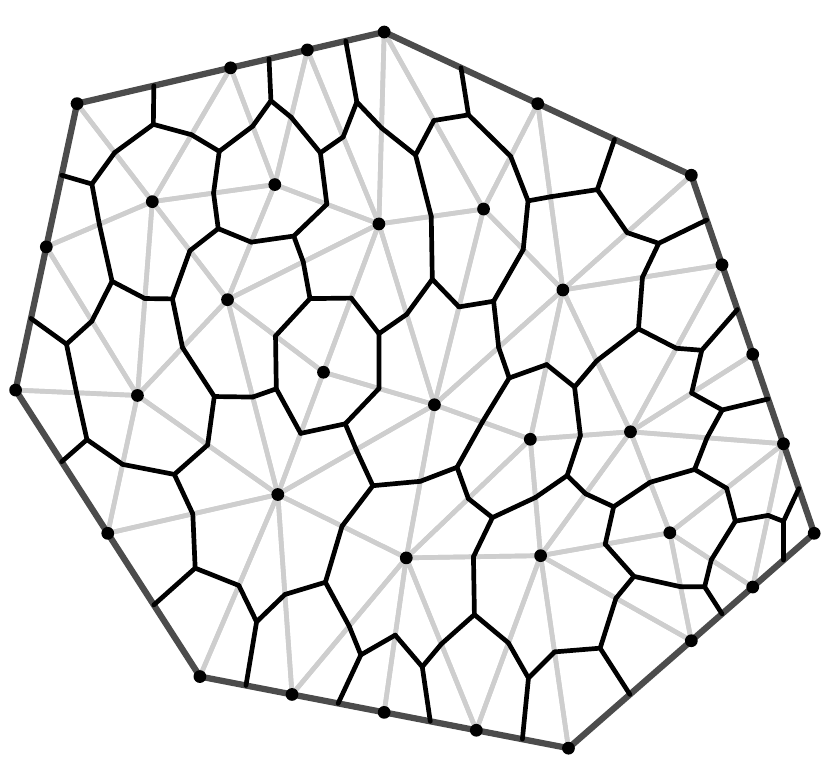}
	\caption{ The dual mesh (black) is constructed by joining the centroids of the primal triangular mesh (grey).  The volumes of these dual cells define the weights of  an integration scheme based at the nodes of the primal mesh.\label{control_volume}}
\end{figure}

In order to complete the model specification, we must define an integration scheme to be used in \eqref{approx_likelihood}. The simplest option is to attach to each node in the mesh a
region $V_i$ for which the value of the basis function $\phi_i(s)$ is
greater than the value of any other basis function.  This
construction, shown in Fig.~\ref{control_volume}, corresponds to the
important notion of the dual mesh.  The corresponding
integration rule sets $\tilde{s}_i$ to be the node location and
$\tilde{\alpha}_i = |V_i|$ to be the volume of the dual cell. This approximation, known as the midpoint rule, is
second-order accurate on a regular grid but will be first-order accurate on an irregular mesh. We can use the structure of the
 mesh in other ways when constructing the integrator,  for example constructing an integration scheme as the sum of
optimal Gaussian integration rules on the individual triangles in the mesh.
The weights and integration points for general triangles can be found in books on numerical analysis or finite
element methods \citep{ErnGuermond}.  We  discuss this further in the Appendix.

\section{Convergence of the approximations} \label{sec:convergence_short}

The proposed method \eqref{approx_likelihood} for approximating the likelihood in a log-Gaussian Cox process has two distinct approximations: one to the integral in the likelihood and another to the latent Gaussian random field.  Here, we show that both of these  converge.  The proofs and more general statements of all of the results can be found in the Appendix.

The first aspect of the approximation, discussed in Section \ref{sec:approx}, replaces the intractable integral of the random intensity with a numerical quadrature scheme.  The following theorem  shows that, for any Gaussian random field $Z(\cdot)$, this approximation converges and the Hellinger distance between the true posterior distribution and the posterior distribution constructed from the approximate likelihood is bounded by the error in the integration scheme.

\begin{theorem} \label{thm:informal_integration}
Assume that $Z(\cdot)$ has, almost surely, $k$ square-integrable derivatives, and that the $p$-point integration scheme in \eqref{approx_likelihood} has deterministic error of order $p^{-k}$. Then the Hellinger distance between the posterior distributions generated with the true and approximate likelihoods is $\mathcal{O}(p^{-k})$.
\end{theorem}

An interesting aspect of using stochastic partial differential equation models as our finite-dimensional Gaussian random field is that the prior distribution converges as the mesh is refined \citep{Lindgren2011,simpson2012think}.  This is distinct from  predictive processes or fixed-rank kriging approaches, where the finite-dimensional model \eqref{basis_expansion} is taken as the true underlying model.  The convergence of this approximation was established by \citet{Lindgren2011}. The following theorem refines this result and shows that the approximate posterior distribution converges weakly.  

\begin{theorem}
Assume that the observation window is a convex polygon in $\mathbb{R}^2$ and let $h$ be the maximum  edge length in the mesh.  Let $Z(\cdot)$ be the solution of \eqref{SPDE} with $\alpha=2$. 
If $G(\cdot)$ is a uniformly Lipschitz continuous function, then the error in the posterior expectation  ${E}\{G(Z) \mid y \}$ due to the stochastic partial differential equation approximation is of order  $h^{1-\epsilon}$ for any  $\epsilon>0$.
\end{theorem}



\section{Examples} \label{sec:examples}
\subsection{Log-Gaussian Cox processes with extensions}
We  consider the application of log-Gaussian Cox processes in three increasingly complicated situations.  In the first, a log-Gaussian Cox process with covariates is fitted to a real data set observed everywhere in a rectangular area \citep{art451, illianalb:11}. The second example is a simulation study in the vein of \citet{Chakraborty2011}, where the point pattern is incompletely observed due to varying sampling effort across the region of interest.  The third case-study is inspired by the problem of mapping the risk associated with freak waves on oceans. We have constructed a  point process defined only on the world's oceans, i.e.,\  over a very irregular, multiply-connected bounded region on a sphere. To the best of our knowledge, no other method can be practically extended to fit a log-Gaussian Cox process in this situation. 

The examples are run using the \texttt{R-INLA} package \citep{art451,martins2013bayesian} which implements both the stochastic partial differential equation models and the integrated nested Laplace approximation in the statistical computing language \texttt{R} \citep{RManual}.  
Wherever not specified otherwise, we use independent Gaussian prior distributions with mean $0$ and variance $100$ on  $\log\kappa$ and $\log\tau$. In all examples, $\alpha=2$.

\subsection{Comparison with a lattice-based approach for rainforest data}\label{seq:rainforest}

\begin{figure}[t]
\subfigure[]{
  \centering
	\includegraphics[width=0.45\textwidth]{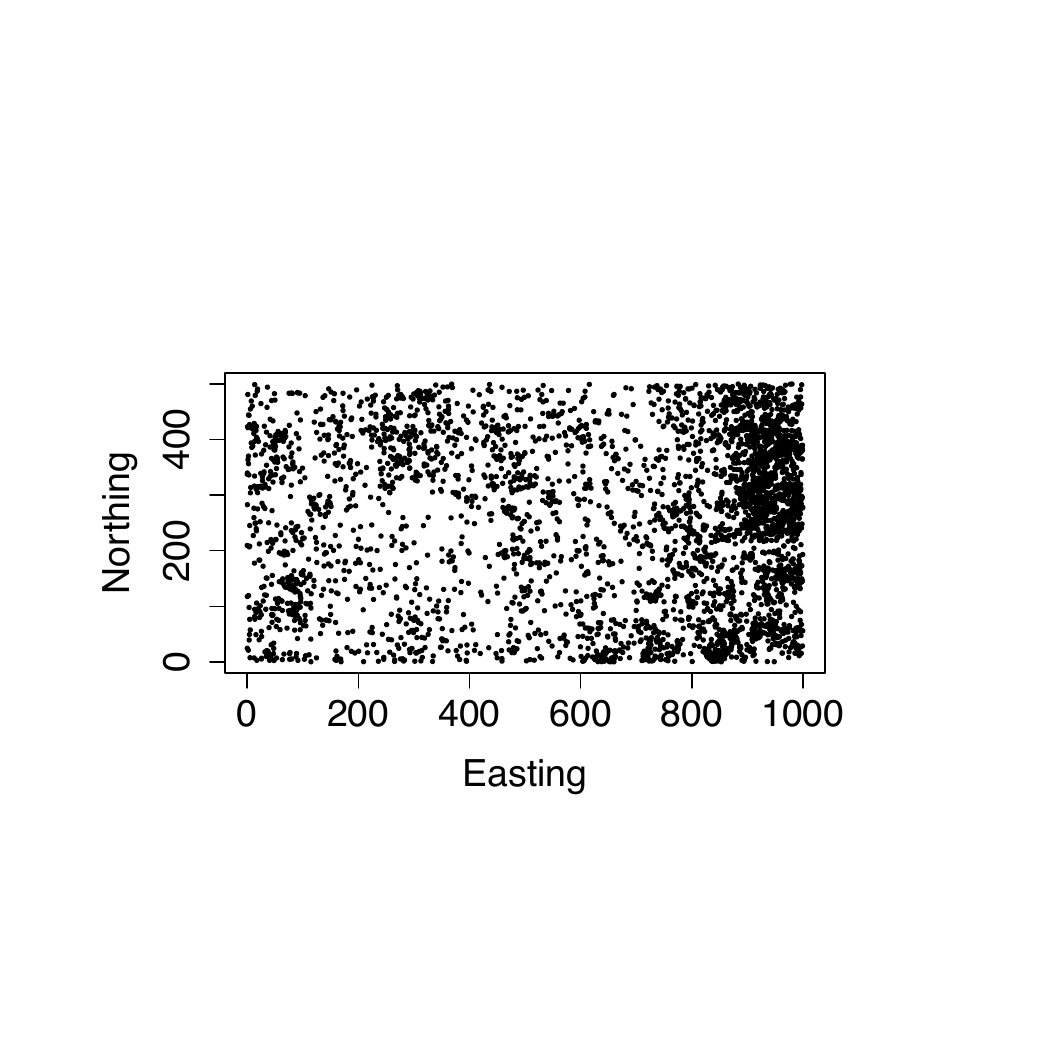}
	\label{rain.dat}
}
\subfigure[]{
\centering
\includegraphics[width=0.45\textwidth]{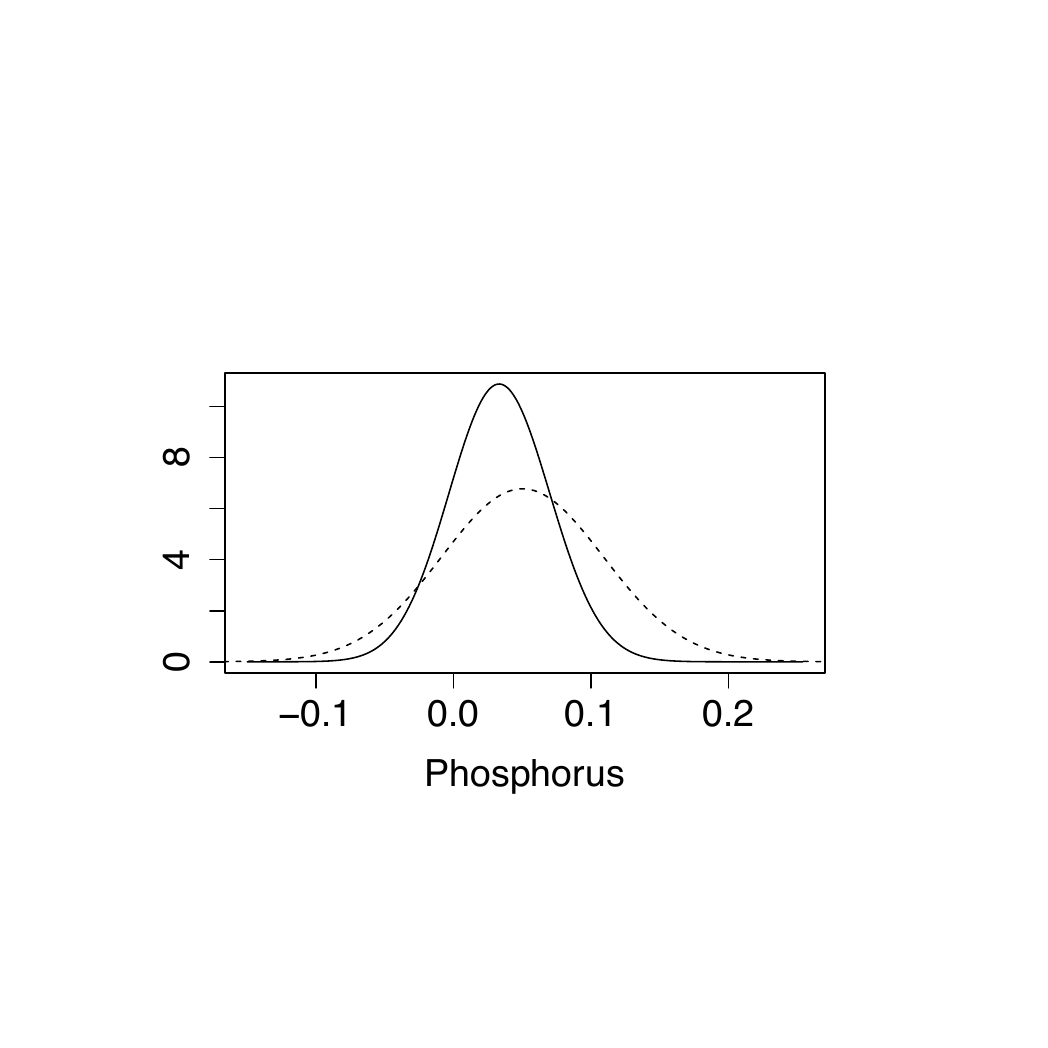}
\label{rain.P}
}
\caption{The effect of soil potassium levels on the location of \textit{Protium tenuifolium}. (a) The location of  \textit{Protium tenuifolium};  (b) The posterior covariate effect of phosphorus, using the standard lattice method (dashed), and the   stochastic partial differential equation approach (solid). \label{forest1}} 
\end{figure}

This case-study is a standard application of spatial point processes, associating species with soil properties in tropical rainforests.  The complete data set consists of the location  of all trees with diameter at breast height of 1cm or greater, for a total of $319$ tree species within a 50 ha rainforest plot on Barro Colorado Island in Panama that has never been logged.  We model the large spatial pattern formed by 4294  trees of species \textit{Protium tenuifolium}, shown in Fig.~\ref{rain.dat}, relative to the covariate phosphorus, which is given on an interpolated grid 
\citep{John2007,hubbellal:99,condit:98}.  
The plot on Barro Colorado Island is only one plot within a large network of 50 ha plots t established as part of an international effort to understand species survival and coexistence in species-rich ecosystems \citep{burslemal:01}.

Data sets with a similar structure have  been analysed both with descriptive \citep{lawal:09} and model-based approaches \citep{art399, waagepetersenal:09, wiegandal:07}.  Integrated nested Laplace approximation can be used to fit a log-Gaussian Cox process to similar data, and also to a joint model of the pattern and covariates \citep{art451,illianalb:11}.   For illustration, we fit a simple model, where the latent field is  $
Z(s) = \mu + \beta P(s)  + x(s)
$, where $\mu$ is a constant mean, $P(s)$ is a spatially varying covariate describing the level of  phosphorus in the soil and $x(s)$ is an approximately intrinsic stochastic partial differential equation model with $\kappa = 0.0014$, which corresponds to a range much larger than the spatial domain.  
The parameter $\log(\tau)$ is assigned a vague Gaussian prior distribution with mean zero and variance $1000$.

\begin{figure}[t]
\subfigure[]{
  \begin{minipage}[b]{0.45\textwidth}%
  \centering%
~\vspace*{-20mm}\\%
	{\includegraphics[width=\linewidth]{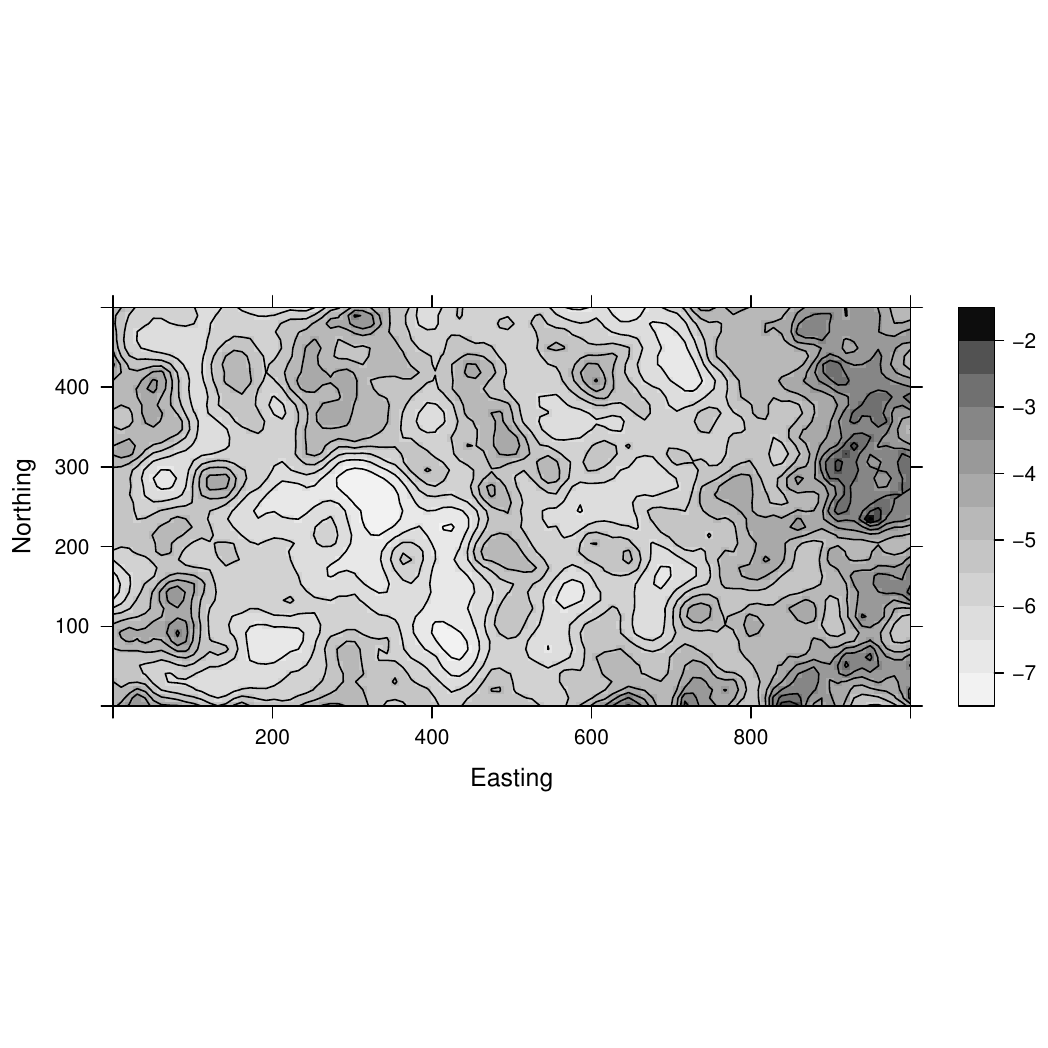}}%
\vspace*{-15mm}%
  \end{minipage}
  \label{lattice.rain}
}
\subfigure[]{
  \begin{minipage}[b]{0.45\textwidth}%
  \centering%
  ~\vspace*{-20mm}\\%
    {\includegraphics[width=\linewidth]{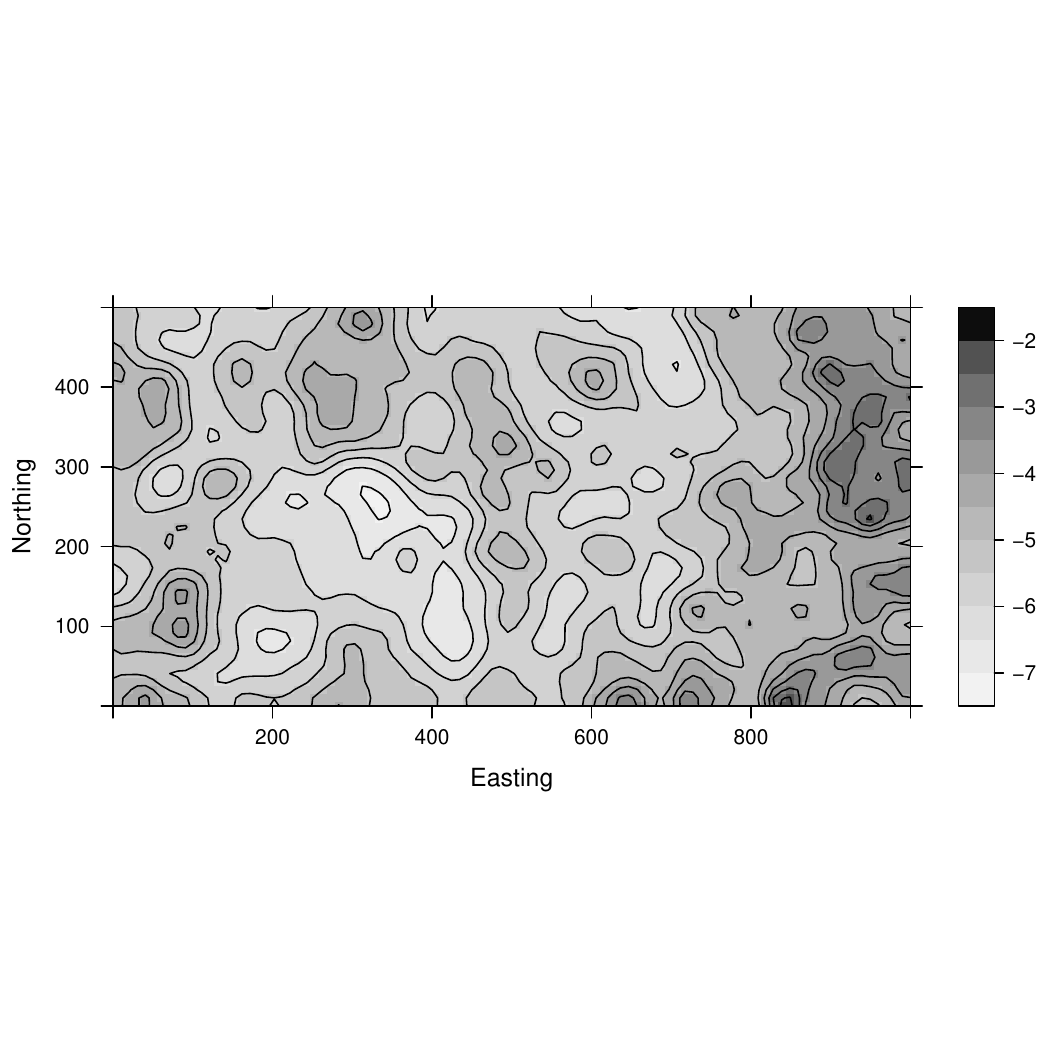}}\\%
    \vspace*{-15mm}%
  \end{minipage}
\label{spde.rain}
}
\caption{Estimated spatial effects for \textit{Protium tenuifolium}: (a) Using a standard lattice point process model; (b) Using the  stochastic partial differential equation approach. \label{forest:means}}
\end{figure}

 For comparison, we fit a lattice model with linear predictor $
 z =  \mu 1 +  \beta P + x,
$ where $1$ is a vector of ones, $P$ the phosphorus concentration, $x\sim N(0,\tau^{-1}Q^{-1})$ is an intrinsic second-order conditional autoregression  \citep{book80} and $\tau \sim Ga(1,10^{-5})$.   Both   models required around $25$ seconds to fit in \texttt{R-INLA}.  The posterior means for the spatial random effects are shown in Fig.~\ref{forest:means} and are centred at the same location.  We believe the difference between the posterior distributions can be accounted for by the different prior distributions for $x$ and the different precision parameters. The posterior distribution for the effect of soil phosphorus  on the locations of trees are shown in Fig.~\ref{rain.P}.

\subsection{Incorporating variable sampling
  effort} \label{sec:sampling}

A major challenge when applying spatial point process models
to real data sets is that the point pattern is  rarely captured
exactly, so sampling effort must be included in the observation
process \citep{chakraborty2010analyzing,Chakraborty2011, niemi10}.  In this example, we  consider the case where there is a sub-area in the data set with no measurements,
but where presences are possible.  This type of situation
occurs, for instance, when considering the spatial distribution of an
animal species over an area that contains a region that is impossible
to survey  \citep{elithal:06}.  In a related situation, data sampling effort varies spatially and is higher in areas where the scientists expect a good chance of presence, as in preferential sampling models \citep{diggle2010geostatistical}.

\begin{figure}[t]
\subfigure[]{
  \centering
	\includegraphics[width=0.45\textwidth,valign=t]{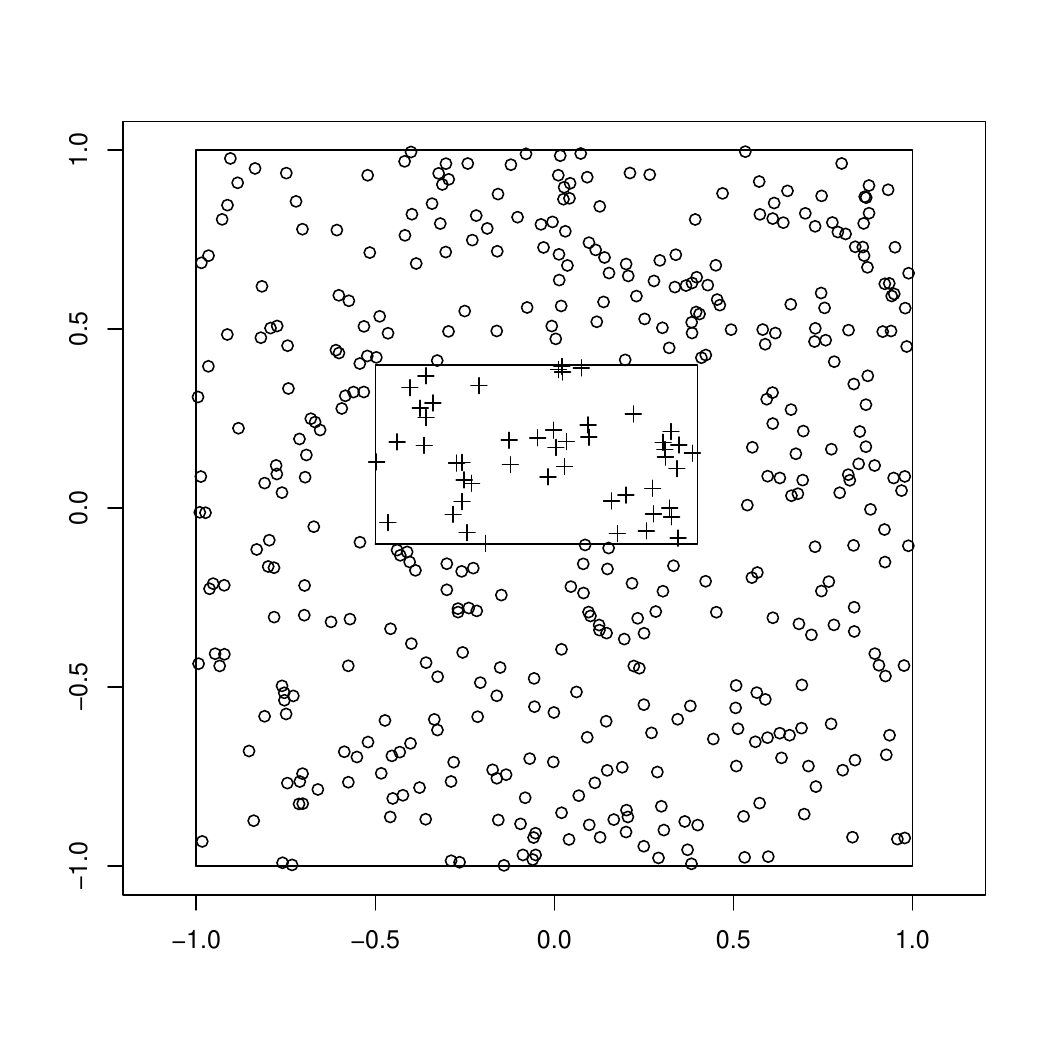}	
	\label{hole_data}
}
\subfigure[]{
\centering
\includegraphics[width=0.46\textwidth,valign=t]{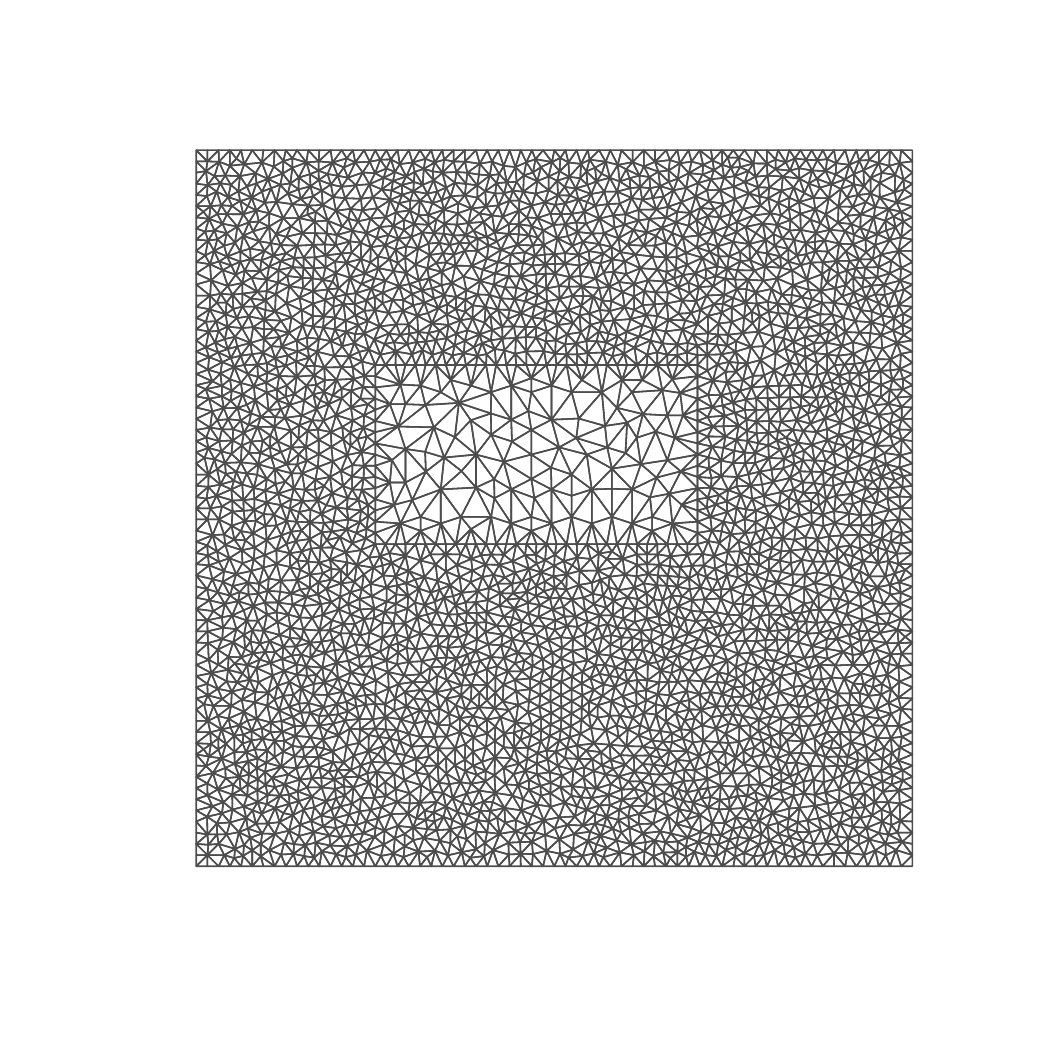}
\label{hole_mesh}
}
\caption{Simulated data with a hole in the sampling
          effort. (a) The inner rectangle borders the area in which there
          was no sampling, and the plusses show the points that
          were missed due to incomplete sampling; (b)  A mesh that
          takes into account the lack of sampling effort in the
          rectangular region.  \label{holes}} 
\end{figure}

Following \citet{Chakraborty2011}, we include known sampling effort
in our model by writing the intensity as $
\lambda(s) = S(s) \exp\{Z(s)\}
$, where $S(s)$ is a known function describing the sampling effort at
location $s$.  In this example, we assume that the point pattern
has been observed perfectly, except in a rectangle  where the pattern is not observed; see Fig.~\ref{hole_data}. We therefore define
$S(s)$ to be zero inside this rectangle and unity everywhere else. It is
straightforward to see from \eqref{likelihood} that, with this choice
of $S(s)$, the unsampled area does not contribute to the integral in
the likelihood.  We can therefore choose the mesh to be quite coarse in
this area, as long as this does not adversely affect the stochastic partial differential equation
approximation to the random field.  Figure \ref{hole_mesh} shows a
mesh that has been coarsened in a rectangular region corresponding to
a hole in the sampling effort.  When coarsening the mesh, it is important to remember that we still want small triangles in the vicinity of the observed region, and we want these to gradually change to larger triangles. This ensures that the  stochastic partial differential equation approximation is stable.  In Fig.~\ref{hole_mesh} this transition can be clearly seen.   The changes to the \texttt{R-INLA} code necessary to add sampling effort to
basic point process code are minimal. This method can be extended in a straightforward
manner to cover more complicated designs, although \citet{Chakraborty2011} suggest it is necessary to assume that the design is known. 

\begin{figure}[ht]
\subfigure[]{
  \centering
	\includegraphics[width=0.45\textwidth]{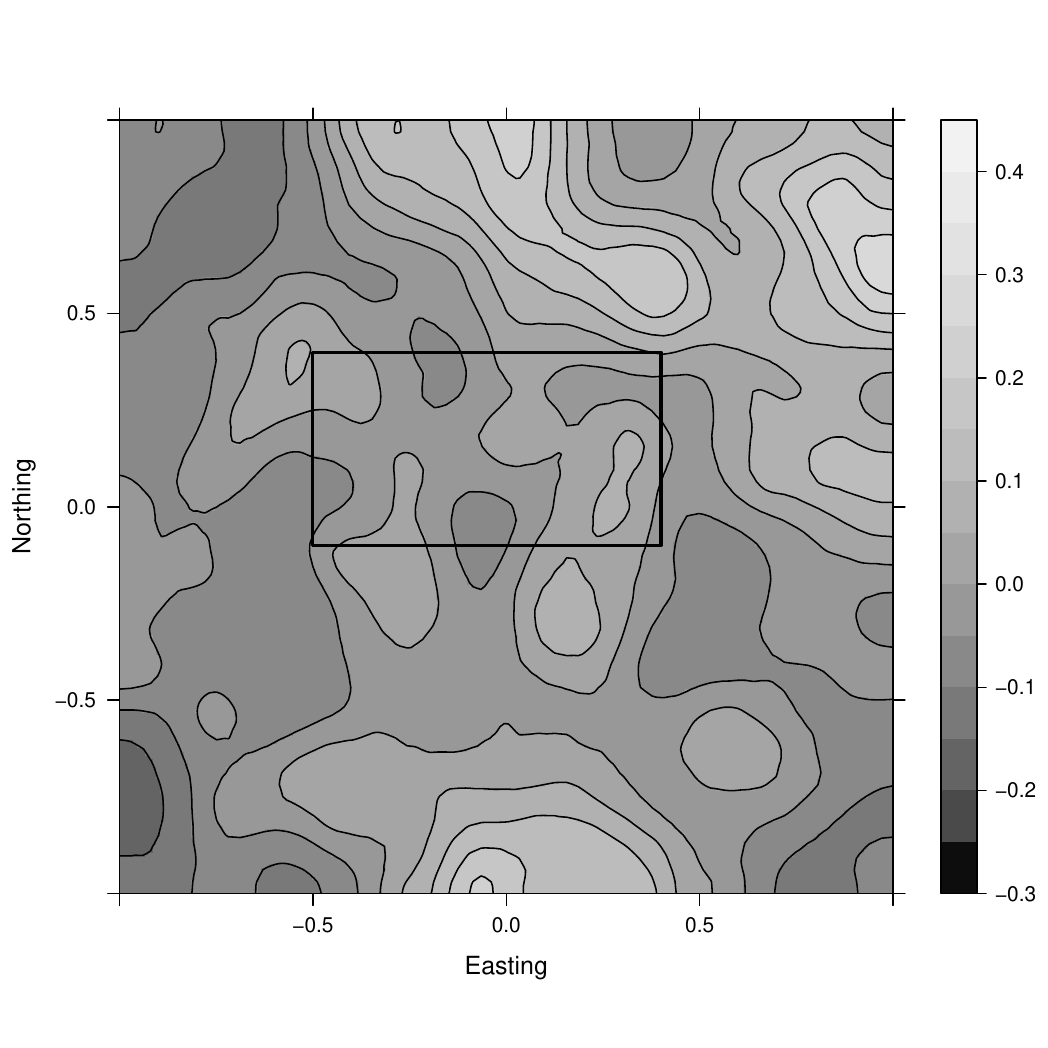}
	\label{complete_data}
}
\subfigure[]{
\centering
\includegraphics[width=0.45\textwidth]{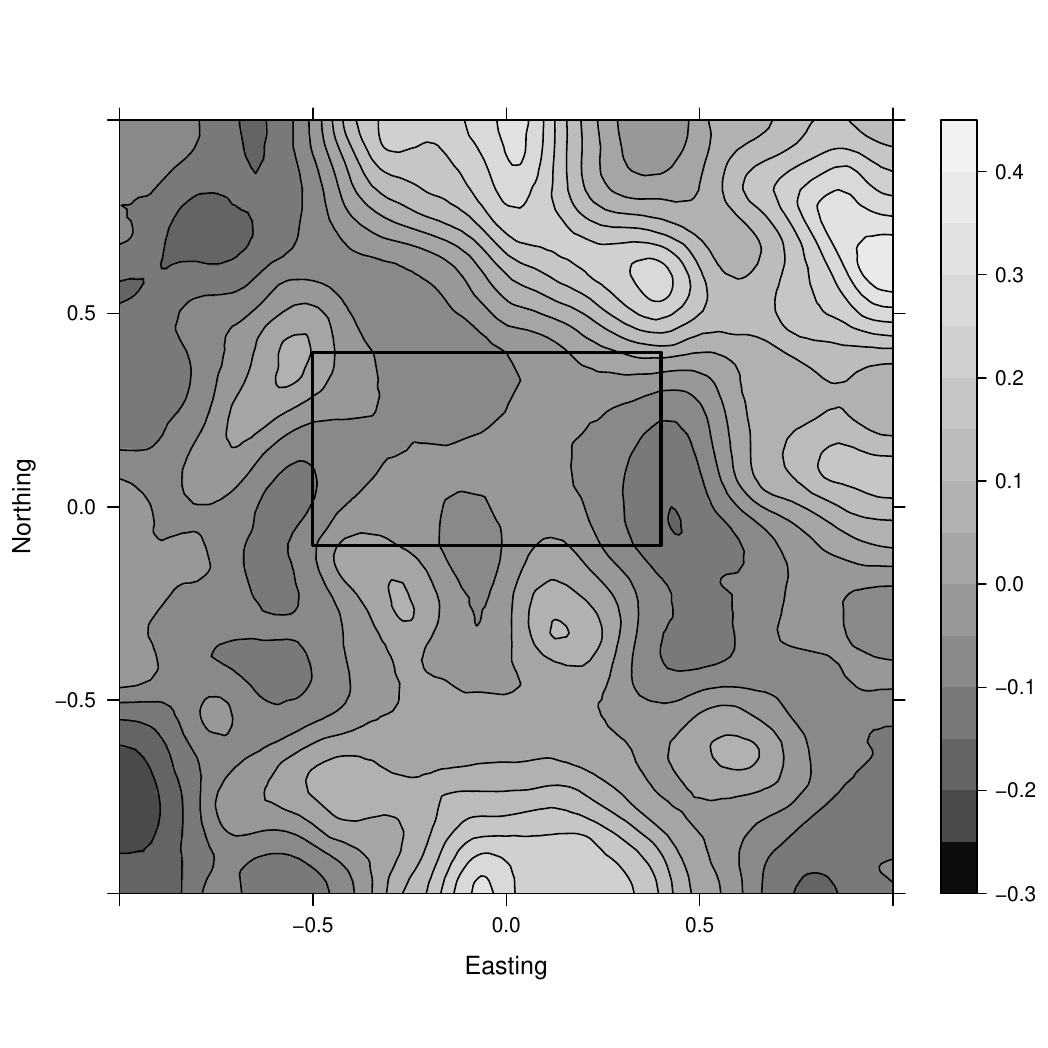}
\label{incomplete_data}
}
\caption{The posterior mean of the spatial effect for variable sampling effort (Section~\ref{sec:sampling}): (a) Using the complete simulated point pattern; (b) Using the incomplete, partially observed point pattern.  The large scale features of both fields are  similar in areas in which the point pattern was sampled.\label{holes_mean}}
\end{figure}

In order to test our method on this type of problem, we simulated a log-Gaussian Cox process on $[-1,1] \times [-1,1]$ and  removed the points from the rectangle $[-0.5,0.4] \times [-0.1,0.4]$ to simulate the variable sampling. The simulated data set is shown in Fig.~\ref{holes}, and the difference in the posterior mean generated from the full data and the censored data is shown in Fig.~\ref{holes_mean}.  There is very little difference between the two posterior means outside  the censored area, whereas there are missing features within the censored area.  We also compared the results obtained for two different meshes with the same maximum edge length, a regular lattice that covers the entire domain and contains $4225$ points, and the irregular mesh consisting of $3850$ points that is coarsened in the censored area, shown in Fig.~\ref{hole_mesh}.  The posterior marginal distributions for the parameters for these two meshes are extremely similar.  In order to avoid boundary effects in the finite-dimensional  random field,  it is important to have  points inside the censored area to ensure that the random field behaves properly. 
Use of the mesh  correctly adapted to the problem  resulted in a significant decrease in computational time.  With the regular grid, the full inference took $37$ seconds on a Linux laptop with a 2.2 GHz i7 4702HQ processor, whereas the computation on the irregular mesh required only $24$ seconds, a $35\%$ reduction by coarsening $11\%$ of the total mesh.

%
%
%

\subsection{A point process over the ocean} \label{sec:ocean}
  
In applications, point processes often occur over complicated domains rather than rectangles, and the topology, topography and geometry of the domain will typically be meaningful  when modelling the covariance structure, see the discussion of \citet{art457} in the context of spatial smoothers.  For this case-study, we have simulated a log-Gaussian Cox process on the oceans, motivated by a model for assessing the risk of freak waves. 

The oceans form a non-convex, multiply-connected bounded region on the sphere and it is, therefore, necessary to construct a Gaussian random field model over this region.  The main complication beyond those considered by \citet{Lindgren2011} is that we need a model for the covariance at the boundary.  This difficult issue has been discussed very little in the statistics literature.  As we are working with simulated data, we can choose a relatively simple, yet realistic, boundary model.  
We expect that wave heights vary more near the coast than in the deep ocean and, as the designation of a freak wave is relative to the expected wave height, the random field is defined using the Neumann boundary conditions, which approximately doubles the variance near the boundaries of the domain \citep[ Theorem 1]{Lindgren2011}.  

\begin{figure}[t]
\subfigure[]{
  \begin{minipage}[b]{0.45\textwidth}%
  \centering%
~\vspace*{-10mm}\\%
	{\includegraphics[width=\linewidth]{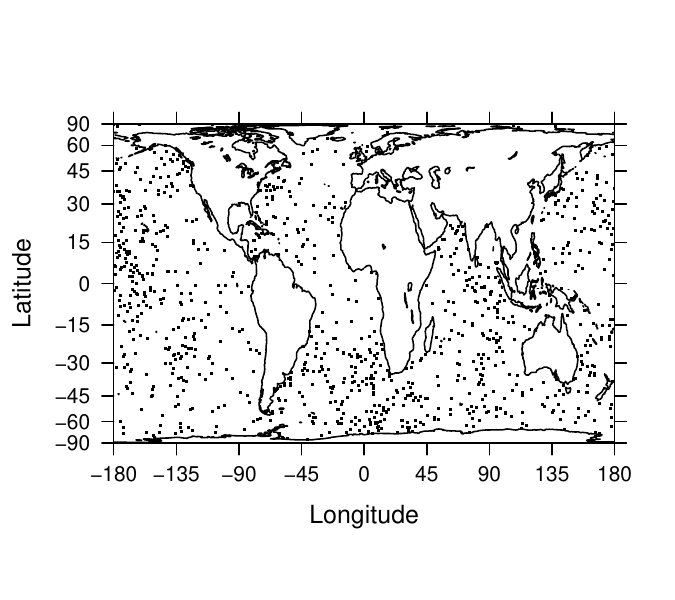}}%
\vspace*{-5mm}%
  \end{minipage}
  \label{ocean_points}
}
\subfigure[]{
  \begin{minipage}[b]{0.45\textwidth}%
    {\includegraphics[width=\linewidth]{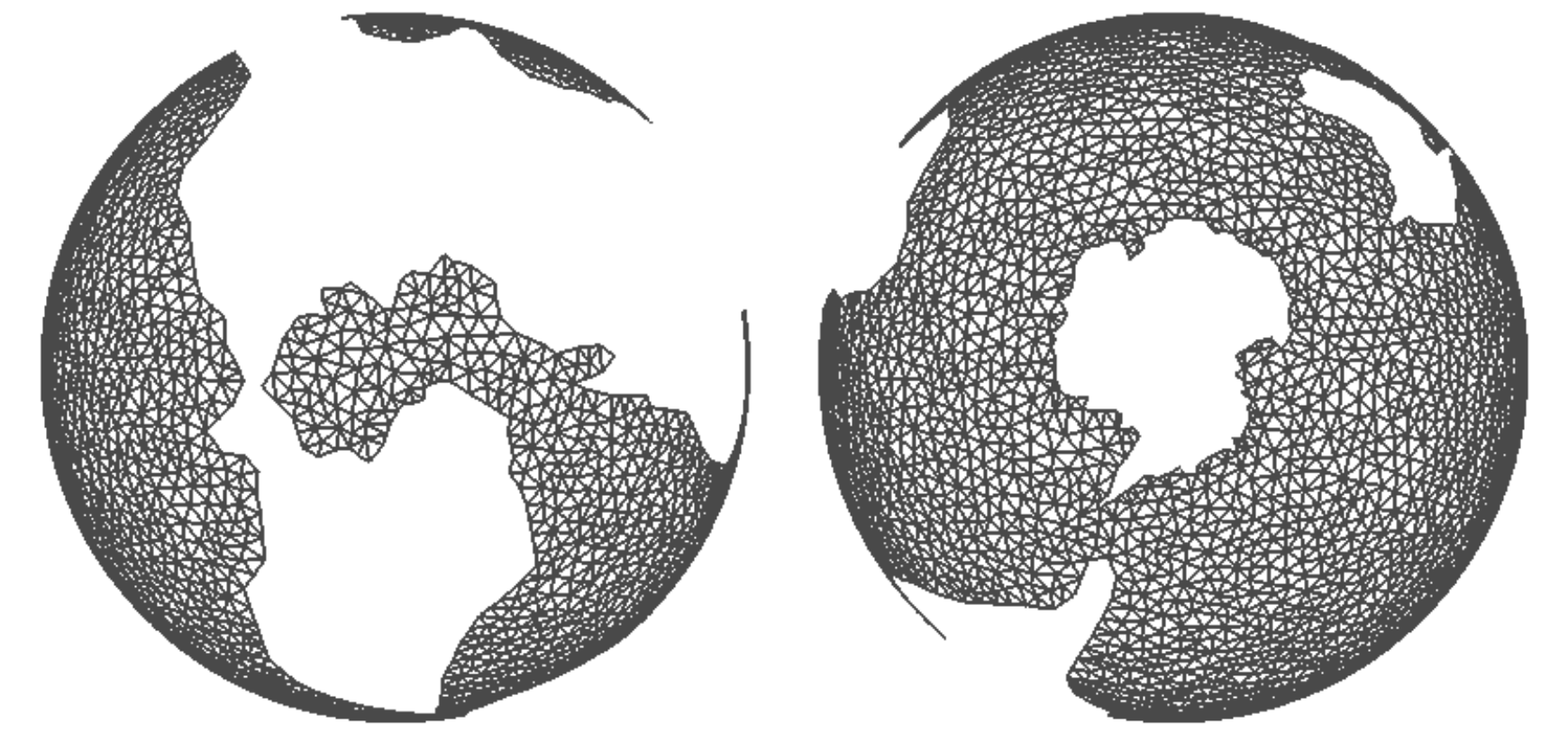}}\\%
    \vspace*{7mm}%
  \end{minipage}
\label{ocean_mesh}
}
\caption{ (a) A simulated log-Gaussian Cox processes over the oceans; (b) A mesh that covers the oceans. \label{ocean_data}}
\end{figure}

\begin{figure}[ht]
\subfigure[]{
  \centering
	\includegraphics[width=0.45\textwidth]{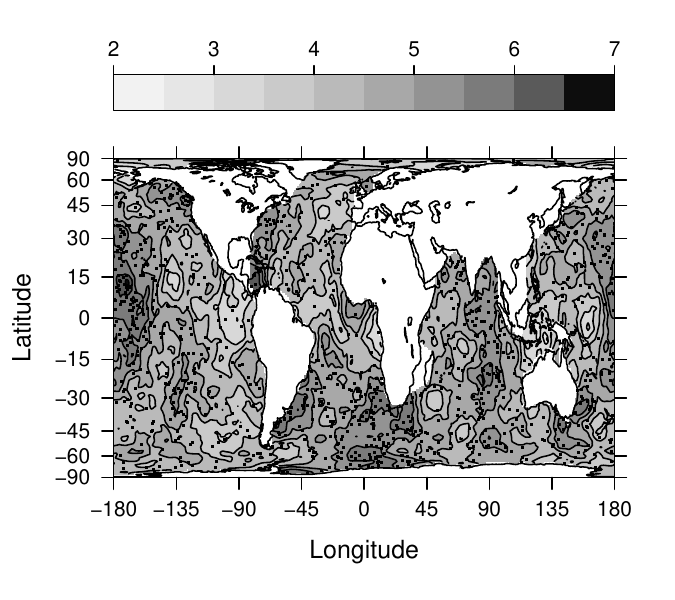}	
	\label{ocean_true}
}
\subfigure[]{
\centering
\includegraphics[width=0.45\textwidth]{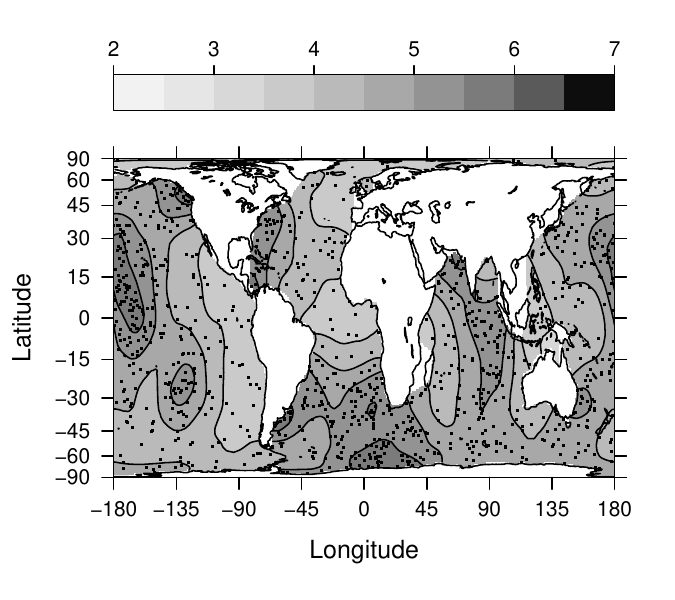}
\label{ocean_mean}
}
\caption{Inference for a point process over the oceans. (a) True surface from the latent Gaussian random field used to generate the sample in Fig.~\ref{ocean_data}; (b) Posterior mean of the latent spatial effect. Note that the large scale behaviour is the same for both figures. \label{ocean_latent}}
\end{figure}

The  point process  shown in Fig.~\ref{ocean_data} was constructed by simulating a Gaussian random field associated with the mesh in Fig.~\ref{ocean_mesh}.  The resulting point pattern has 913 points.  Inference was performed on this model and the posterior mean is shown in Fig.~\ref{ocean_mean}.  The posterior mean shows the same large-scale features as the sample used to generate the log-Gaussian Cox process, see Fig.~\ref{ocean_true}, with the expected loss of information due to the uninformative nature of point pattern data.

\begin{figure}[ht]
\subfigure[]{
\centering

\includegraphics[width=0.45\textwidth]{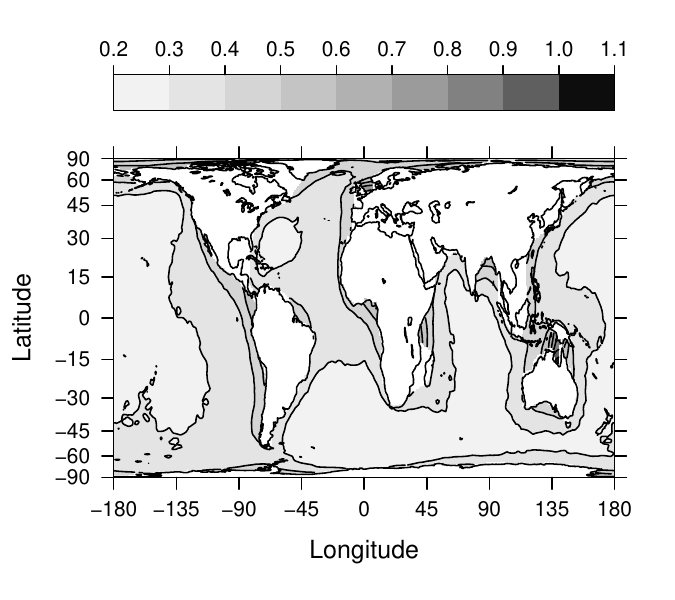}
\label{ocean_sd}
}
\subfigure[]{
  \centering
	\includegraphics[width=0.45\textwidth]{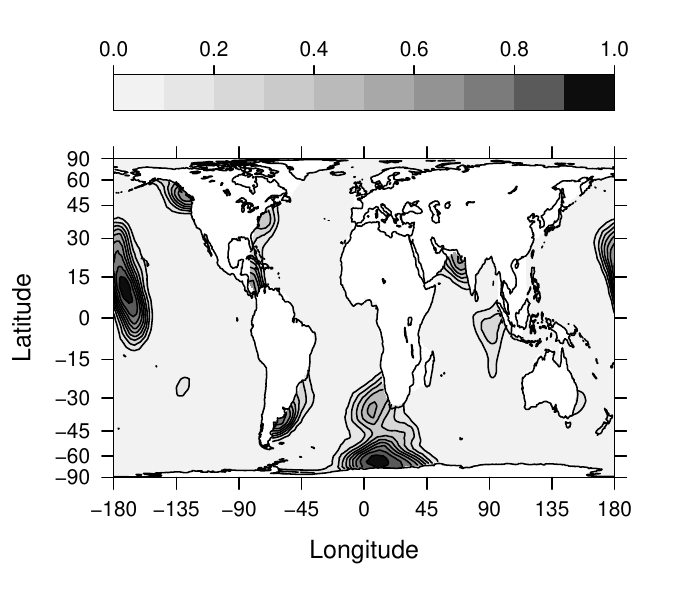}
	
	\label{risk_map}
}

\caption{Inference for a point process over the oceans. (a) The pointwise posterior standard deviation for the log risk surface; (b) The posterior risk map $\rm{pr}\log\lambda(s) > 5.5 \mid y\}$. \label{ocean_misc}}

\end{figure}

Effects induced by the boundary conditions can be seen in Fig.~\ref{ocean_misc}.  The pointwise posterior standard deviation of the latent Gaussian field is shown in Fig.~\ref{ocean_sd}.   The standard deviation is reasonably constant away from the coasts, but is much higher near the boundaries.  There are some interesting effects in the Gulf of Carpentaria in Australia, and  in the North Sea.  This is  an effect of the prior model, which increases the variance near the boundaries and in areas with high curvature of the coastline.   

In the context of freak wave modelling, the most important result is displayed in Fig.~\ref{risk_map}, showing the probability that the log-risk will be greater than $5.5$. Once again we see pronounced effects near the coastlines. This type of map can  easily be computed using  the function inla.pmarginal  in the \texttt{R-INLA} package.   It is also  possible to use the \texttt{excursions} package \citep{bolin2012excursion} in \texttt{R} to construct joint exceedance maps.

\section{Discussion and future work}
The approximation to analyse log-Gaussian Cox processes introduced in this paper is valid also when using kernel methods \citep{art473}, predictive processes \citep{art444} or fixed-rank kriging \citep{art445}.  The problem with using these  methods in the given context is that their basis functions are typically non-local and, therefore, the point evaluation matrices $A_i$ in \eqref{final_likelihood} are dense; see \citet{simpson2012think} for a further discussion of the choice of basis functions in spatial statistics. 


In Section \ref{sec:ocean}, we consider a point process over a complicated region of the sphere.  To the best of our knowledge, there are no other applicable inference methods for this example that include a covariance model at the boundaries. In general, modelling of boundary effects for point processes has not previously been discussed in the literature. We argue that by using Neumann, or no-flux, boundary conditions the variance at the boundaries increases.  Similarly, Dirichlet boundary conditions, which correspond to fixing the value of the field at the boundaries, decrease the variance. A future question is to construct good boundary models, and study their effect  in a statistical context. 

There is work to be done on the theoretical properties of the approximation presented in this paper.  Some partial results are given in the Appendix, but they are not the complete story.  In particular, it would be  interesting to study the effect of both the likelihood approximation and the finite-dimensional approximation of the hyper-parameters of the model.  These parameters, which control range, variance and, in more complicated cases, non-stationarity, are often of scientific interest and determining the rate of convergence will help us understand their interpretations.

Moving our considerations to more general finite-dimensional expansion \eqref{basis_expansion}, it is also of  interest to quantify the link between the basis functions $\phi_i(s)$ and the statistical properties of the estimator.  Although there has been  work by  \citet{stein2013limitations}, there are a number of open questions.  This is a challenging problem as the interest is in non-asymptotic behaviour both in the number of basis functions and in the amount of data.  In order to do practical spatial statistics, we need to give something up and often methods will be asymptotically incorrect. However, it may be that in realistic regimes, the resulting statistical error  is manageable.

Finally,  the approximation in Section \ref{sec:approx} applies even when the latent random field $Z(s)$ is not Gaussian.  The only requirement is that it has the basis function expansion \eqref{basis_expansion} and that the statistical properties of $z$ are known.  In particular, this approximation applies to stochastic partial differential equation models with non-Gaussian noise.  This has been investigated for type-$G$ L\'{e}vy processes, and especially for Laplace random fields \citep{BolinNonGaussian}. 
Similarly, replacing Gaussian white noise with Poisson noise would result in shot-noise Cox process models of the Mat\'{e}rn type.  It may be possible to avoid the assumptions that the random field is Gaussian in the Appendix.  The main use of Gaussianity is in the form of Fernique's theorem, which is a statement about the tails of a Gaussian random field and it is possible that similar results would hold for non-Gaussian fields after modifying the growth conditions on  the likelihood and the functionals.

\section*{Acknowledgement}
The authors wish to thank the associate editor and the anonymous reviewers for their useful suggestions and, in particular, for pushing us to work out the convergence results. The authors  gratefully acknowledge the financial support of Research Councils UK for Illian.  We would also like to thank David Burslem for long discussions on the rainforest data.

The BCI forest dynamics research project was made possible by National Science Foundation grants to Stephen P. Hubbell:
DEB-0640386, DEB-0425651,DEB-0346488, DEB-0129874, DEB-00753102, DEB-9909347, DEB-9615226, DEB-9615226, DEB-9405933, 
DEB-9221033, DEB-9100058, DEB-8906869, DEB-8605042,DEB-8206992, DEB-7922197, support from the 
Center for Tropical Forest Science, the Smithsonian Tropical Research Institute, the John D. and Catherine T. MacArthur
Foundation, the Mellon Foundation, the Small World Institute Fund, and numerous private individuals, and 
through the hard work of over 100 people from 10 countries over the past two decades. The plot project is part of 
the Center for Tropical Forest Science, a global network of large-scale demographic tree plots.



\appendix
\section*{Appendix} 
\subsection{Likelihood approximation}  \label{appendix:likelihood}

Throughout this appendix, we assume that the parameters in the covariance model for $Z(s)$ are known and fixed.  We show that, for a fixed random field $Z(s)$, the posterior distribution computed using the likelihood approximation converges strongly to the true posterior distribution, and the rate of convergence can increase as the smoothness of the field increases. We also show that for either the true or approximate likelihood, the posterior distribution generated using the finite-dimensional stochastic partial differential equation model converges weakly to that generated using the limiting Mat\'{e}rn model.  Finally, we  show that under  further assumptions  the  convergence rate  depends both on the basis functions and the smoothness of the underlying random field.  The main tools used in this appendix come from the inverse problems literature, surveyed in \citet{stuart2010inverse}, which deals with inference of indirectly observed continuous 
Gaussian random fields.

In order to show that the approximate posterior distributions converge to the true posterior distribution, it is useful to re-write the problem in terms of measures.  Let $\mu_0(A) = \mbox{pr}\{Z(\cdot) \in A\}$ be the Gaussian measure defined by the Gaussian random field prior measure on $Z(\cdot)$.  If we define 
$$ \Phi(Z;Y) = \int_\Omega \exp\{Z(s)\} \,ds  - \sum_{s_i \in Y} Z(s_i),$$ then the posterior probability measure $\mu$ for $Z (\cdot)$ conditioned on $Y$ can be defined through its Radon--Nikodym derivative 
$
\sfrac{d\mu}{d\mu_0}(Z) \propto M^{-1}\exp\{-\Phi(Z;Y)\},
$
where $M$ is a normalising constant required to ensure that $\mu$ is a probability measure. We can, in a similar fashion, define the approximate posterior measure as 
\begin{equation} \label{eqn:approx_posterior}
\frac{d\mu^p}{d\mu_0}(Z) \propto M_p^{-1} \exp\{-\Phi^p(Z;Y)\},
\end{equation}
where 
$\Phi^p(Z;Y) =\sum_{i=1}^p \tilde{\alpha}_i \exp\{Z(\tilde{s}_i)\} - \sum_{s_i \in Y} Z(s_i),$ and $M_p$ is a normalising constant.  \citet{cotter2010approximation} showed that, under conditions $\Phi$ and $\Phi^p$, the Hellinger distance
$$
d_\text{Hell}\left(\mu,\mu^p\right) =\left[\frac{1}{2}\int \left\{ \left({\frac{d\mu}{d\mu_0}}\right)^{1/2} - \left({\frac{d\mu^p}{d\mu_0}}\right)^{1/2} \right\}^2\,d\mu_0\right]^{1/2}
$$
 between the approximate and true posterior distributions converges to zero.    \citet{stuart2010inverse} notes that convergence in Hellinger distance implies convergence in the total variation metric and it can be related to convergence of functionals using the identity 
\begin{equation} \label{eqn:hellinger_weak}
\left\lvert E_{\mu} \{f(Z)\} - E_{\mu'}\{f(Z')\}\right\rvert \leq 2\left[E_{\mu} \{|f(Z)|^2\} - E_{ \mu'}\{|f(Z')|^2\}\right]d_\text{Hell}(\mu,\mu').
\end{equation}
The following theorem shows that their theory applies to our approximate likelihood.


\begin{theorem} \label{thm:likelihood_convergence}
Consider a Gaussian random field $Z(\cdot)$ defined on a Lipschitz domain $\Omega$ and assume that its paths are  almost surely in the Sobolev space $H^\alpha(\Omega)$ with $\alpha > d/2$.  Assume that the integration rule satisfies 
\begin{equation} \label{eqn:quad}
\left| \int_{\Omega} f(s)\,ds - \sum_{i=1}^p \tilde{\alpha}_i f(\tilde{s}_i) \right| \leq C \psi(p) \norm{f}_{H^\gamma},
\end{equation} where $\psi(p) \rightarrow 0$ as $p \rightarrow \infty$ and  $\gamma \leq \alpha$.  Then, as $p \rightarrow \infty$, $
d_\text{Hell}\left(\mu,\mu^p\right) \rightarrow 0$. Furthermore, if $\gamma $ is an integer, then $
d_\text{Hell}\left(\mu,\mu^p\right)  \leq C \psi(p).
$
\end{theorem}

\begin{proof}
  This result follows directly from Theorem 2.4 of \citet{cotter2010approximation} if we can show that the potential is bounded above and below and that the error in the likelihood approximation is integrable.
  Let $\norm{Z(\cdot)}_\infty=\sup_{s\in\Omega}|Z(s)|$, and let $\norm{Y}$ be the number of points in the point pattern $Y$.
  Firstly we note that, by assumption, $\norm{Z(\cdot)}_\infty$ and $\norm{Y}$ are almost surely finite.
  Then, if $\max(\norm{Z}_\infty, \norm{Y}) < r$, straightforward calculation shows that $
\Phi(Z;Y) \leq |\Omega| e^r+ r^2.
$
Similarly, when $\norm{Y} < r$,
\begin{align*}
\Phi(Z;Y) = \int_\Omega \exp\{Z(s)\} \,ds  - \sum_{s_i \in Y} Z(s_i) 
\geq - r \norm{Z}_\infty \geq - Cr \norm{Z}_{H^\gamma},
\end{align*}
where the last inequality follows from Sobolev's embedding theorem and is true for every $\gamma>d/2$.
Similar arguments show that $\Phi^p(Z;Y)$ is also bounded above and below independently of $p$.

 To show that the error in the likelihood induces a similar error in the posterior distribution, we need to verify that, for sufficiently small $\epsilon>0$, there exists a $K >0$ that does not depend on $Z$ such that $
\left | \Phi(Z;Y) - \Phi^p(Z;Y) \right | \leq K \exp\left(\epsilon \norm{Z}_\infty^2\right) \psi(p).
$ By assumption, this reduces to showing that $\norm{\exp\{Z(\cdot)\}}_{H^\gamma} \leq K \exp(\epsilon\norm{Z}_\infty^2)$
for large enough $Z$.  Let $\gamma$ be an integer.  Now, for any realisation of $Z(\cdot): D \rightarrow \mathbb{R}$, there exists an extension $IZ(\cdot):\mathbb{R}^d \rightarrow \mathbb{R}$ such that  $IZ(\cdot) \in H^\gamma(\mathbb{R}^d)$ has compact support and $\left. IZ\right\rvert_D(\cdot) = Z(\cdot)$.  Using the quotient space structure of a Sobolev space on a domain, it follows that 
\begin{align*}
\norm{\exp\{Z(\cdot)\}}_{H^{\gamma}(\Omega)} &= \inf_{H^\gamma(\mathbb{R}^d) \ni \tilde{Z}(s)=Z(s), \; a.s.\; s\in D}\norm{\exp\{\tilde{Z}(\cdot)\}}_{H^\gamma(\mathbb{R}^d)} \\
&\leq C \exp(\norm{IZ(\cdot)}_{L^\infty(\mathbb{R}^d)})\left(\norm{IZ(\cdot)}_{H^\gamma(\mathbb{R}^d)} + \norm{IZ(\cdot)}_{H^\gamma(\mathbb{R}^d)}^\gamma  \right) \\
&\leq C\exp(C\norm{Z(\cdot)}_{\infty}) \left(\norm{Z(\cdot)}_{H^\gamma(\Omega)} + \norm{Z(\cdot)}_{H^\gamma(\Omega)}^\gamma  \right) ,
\end{align*}
where the first inequality follows from Theorems 2 and 3 of \citet{bourdaud2011composition}, the second inequality follows from the boundedness of the extension operator and the  constant $C$ changes from line to line.

\end{proof}

\begin{remark}\label{remark:gap}
The condition that $\gamma$ is an integer can probably be relaxed, but it is an open question whether $\norm{\exp\{Z(s)\}}_{H^\gamma(\mathbb{R}^d)}$ can be bounded for non-integer $\gamma$ in the same way as  in the integer case.  If this was true, it would suggest the use of integration rules of order $\lceil{\alpha}\rceil$ rather than $\lfloor{\alpha}\rfloor$ and would slightly improve the convergence rate. 
\end{remark}

The techniques used to prove Theorem~\ref{thm:likelihood_convergence} also allow us to give a more informative convergence result for the traditional counting process approximation to the log-Gaussian Cox process than those considered by \citet{waagepetersen2004}. 

\begin{corollary} \label{cor:lattice}
Assume $\alpha \geq 2$. Then the classical $(p+1)\times (p+1)$ lattice approximation to the log-Gaussian Cox process converges in the Hellinger distance at a rate of $\mathcal{O}(p^{-1})$.
\end{corollary}
\begin{proof}
For simplicity, we will assume that the observation window $D$ is  square and the lattice is equally spaced in both directions. The lattice approximation is of the form \eqref{eqn:approx_posterior} with 
\begin{equation} \label{eqn:lattice}
\Phi^p(Z) = \sum_{i,j=1}^p|S_{ij}|\exp\{Z(\tilde{s}_{ij})\} - \sum_{i,j=1}^p \norm{Y \cap S_{ij}} Z(\tilde{s}_{ij}),
\end{equation}
where $S_{ij}$ is the $(i,j)$ lattice cell and $\tilde{s}_{ij}$ is the centroid of $S_{ij}$.   The first term in \eqref{eqn:lattice} is the midpoint rule approximation to $\int_\Omega \exp\{Z(s)\}\,ds$, which, due to the regularity of the lattice satisfies \eqref{eqn:quad} with $\psi(p) = p^{-\gamma}$ $(d/2 < \gamma \leq 2)$, \citep[Theorem 8.5,][]{ErnGuermond}.  The error in the likelihood arising from the approximation of $Z(s_k)$ by $Z(\tilde{s}_{ij})$ for any $s_k \in Y \cap S_{ij}$ can be bounded using Taylor's theorem as $
\left | Z(s_k) - Z(\tilde{s}_{ij}) \right | \leq p^{-1} \sup_{s \in S_{ij}} \sup_{\ell=1,\ldots,d} \left|\frac{\partial Z(s)}{\partial s_{\ell}}\right| \leq Cp^{-1} \norm{Z}_{H^{1+d/2}(S_{ij})},
$
where the second inequality is a consequence of Sobolev's embedding theorem \citep[Corollary 1.4.7]{book107}.  It follows using the arguments in the proof of Theorem \ref{thm:likelihood_convergence} that for a two-dimensional lattice, 
 $
\left | \Phi(Z) - \Phi^p(Z)\right| \leq C\norm{Y} p^{-1}\norm{Z}_{H^2(\Omega)}
$, and the result follows from Theorem 2.4 of \citet{cotter2010approximation}.
\end{proof}

\begin{remark}
Examining the proof of Corollary A\ref{cor:lattice}, it can be seen that the rate of convergence is determined by the binning procedure and using the lattice quadrature rule and the approximate likelihood proposed in this paper the rate of convergence would be $\mathcal{O}(p^{-2})$ for smooth enough fields.
\end{remark}

\subsection{Random field approximation}  \label{appendix:random_field}
While Appendix \ref{appendix:likelihood} shows that for fixed $Z(\cdot)$ the likelihood approximation introduced in this paper converges, this is not enough to show that the posterior distributions computed in Section \ref{sec:examples} converge as we are simultaneously approximating the log-Gaussian Cox process likelihood and the Gaussian random field using the approximation outlined in Section \ref{sec:spde}.  In this appendix, we close this gap when the hyperparameters are fixed and show the convergence of a general class of finite-dimensional approximations to problems in which the indirectly observed unknown random function is equipped \emph{a priori} with a Gaussian random field.

There are a number of technical challenges to showing convergence of this approximation.  The first is that we need to compare a measure on an infinite-dimensional space with a sequence of measures on different finite-dimensional spaces.  We will, therefore, no longer be able to consider convergence in the Hellinger metric, but rather we will consider a weaker mode of convergence of an approximating measure $\nu^{n,p}$ to  $\mu$, that is the convergence of functionals of the form $
\int G(Z_n) \,d\nu^{n,p}(Z_n) \rightarrow \int G(Z)\,d\mu(Z),
$
for  Lipschitz continuous functions that satisfy a growth condition to ensure the functionals are finite.  This is slightly stronger than convergence in distribution, for which bounded Lipschitz functions suffice \citep[Section 8.3]{bogachev2007measure}.  When the finite-dimensional approximation to the Gaussian random field  is computed by truncating its Karhunen--Lo\`{e}ve expansion, \citet{dashti2011uncertainty} showed convergence.  Their techniques, which relied heavily on the idea that truncation of the Karhunen--Lo\`{e}ve expansion is an $L^2(\Omega)$ projection,  are not directly applicable to the approximation outlined in Section \ref{sec:spde}. 


In Section \ref{appendix:finite_dimensional}, we  extend Theorem 2.6 of \citet{dashti2011uncertainty} to a  general class of finite-dimensional approximations. In particular we show that if the approximation $Z_n(\cdot)$ to  $Z(\cdot)$ is stable, in the sense that $\norm{Z_n}_H \leq C \norm{Z}_H$ uniformly in $n$, then the convergence of the functionals is governed by the deterministic error in the pathwise approximation.  In Section \ref{appendix:spde_converge} we show that for approximations of the general form of the stochastic partial differential equation approximation, this error is controlled by the ability of the finite-dimensional basis functions to approximate realisations of the true prior model.  These results mirror previous quantitative results   \citep{Simpson2011, simpson2012think, Bolin2011}, in which the stable, convergent approximation properties of piecewise linear functions were used to argue for the adoption of stochastic partial differential equation models.

\subsection{A general result on the convergence of finite-dimensional approximations} \label{appendix:finite_dimensional}

Let $V \subset H$ be Banach spaces and assume that $\norm{\cdot}_H \leq C \norm{\cdot}_V$. Assume that the Gaussian random field  $Z(\cdot)$ has paths almost surely in $V$ and define the approximate random field $Z_n(\cdot) = R_nZ(\cdot)$, where $R_n:V \rightarrow V_n$ is a deterministic linear operator, and $V_n \subset H$ is an $n$--dimensional vector space that is not necessarily a subspace of $V$.  In the special case that $V_n \subset V$ and $R_n$ is a projector, the arguments of \citet{dashti2011uncertainty} can be used to show convergence.



Extending the notation from Appendix \ref{appendix:likelihood}, we define $\mu_0(\cdot)$ to be the law of $Z(\cdot)$ and consider the infinite-dimensional posterior distribution $\nu(\cdot)$ defined by $
\sfrac{d\mu}{d\mu_0} = M^{-1}\exp\{-\Phi(Z;Y)\}.
$
Similarly, we define the law of $Z_n(\cdot)$ to be $\nu_0^n(\cdot)$ and define the approximate posterior distributions  $\nu^{n,p}$ as $
\sfrac{d\nu^{n,p}}{d\nu_0^n} = M_{n,p}^{-1}\exp\{-\Phi^p(Z_n;Y)\},
$ where $M_{n,p}$ is a normalising constant.  We make the following assumptions on the potential $\Phi(\cdot;Y)$ \citep{dashti2011uncertainty}.

\begin{assumption} \label{assumption:potential}
Consider the potential function $\Phi(\cdot;Y): H \rightarrow \mathbb{R}^+$. Assume that for every $\epsilon >0$ and $r>0$, $\norm{Y} < r$ and  there exists a $C = C(\epsilon,r)>0$, which may change from line to line, such that, for all $Z \in H$, 
$\exp\{-\Phi(Z;Y)\} \leq C \exp(\epsilon \norm{Z}_H^2).$
Also, for every  $Z \in H$, where $\norm{Z}_H < r$, $\Phi(Z;Y) \leq C$ and for every $Z_1, Z_2 \in H$,
$|\Phi(Z_1;Y) - \Phi(Z_2;Y) | \leq C\exp\left\{\epsilon \max(\norm{Z_1}_H^2,\norm{Z_2}_H^2) \right\}\norm{Z_1 - Z_2}_H.$
\end{assumption}

%
%
%

The following theorem says that for nice functionals, the error in the approximation depends on how well the approximate random field $Z_n(\cdot)$ approximates the true random field in a pathwise sense as well as on the quality of the likelihood approximation.  While the argument holds mutatis mutandis for Banach space-valued functionals $G$ \citep[see][]{dashti2011uncertainty}, for  simplicity we restrict ourselves to real-valued functionals.

\begin{theorem} \label{thm:finite_dimensional}
Assume that Assumption \ref{assumption:potential} holds for $\Phi(\cdot;Y)$, $\Phi^p(\cdot;Y)$ and $\Phi^{n,p}(\cdot;Y) = \Phi^p(R_n\cdot;Y)$ uniformly in $n$ and $p$.
Let $G$ be a Lipschitz continuous function such that, for every $\epsilon > 0$, there exists a $C = C(\epsilon) \in (0,\infty)$ such that, for every $Z_1 \in V$ and  $Z_2 \in H$, $
|{G(Z_1) - G(Z_2)}| \leq C\exp\left\{\epsilon \max( \norm{Z_1}^2_V, \norm{Z_2}^2_H )\right\}\norm{Z_1 - Z_2}_H.
$
If the restriction operator $R_n$ satisfies the stability estimate \begin{equation} \label{eqn:stability}
\norm{R_n Z(\cdot)}_H \leq C \norm{Z(\cdot)}_V,
\end{equation}
for all  $Z(\cdot) \in V,$
then $$
e_G = | E_\mu\{G(Z)\} - E_{\nu^{n,p}} \{G(Z_n)\}| \leq C \left\{\sup_{Z(\cdot) \in V}\frac{ \norm{Z(\cdot) - R_nZ(\cdot)}_H}{\norm{Z(\cdot)}_V} + \psi(p)\right\}.
$$
\end{theorem}
\begin{proof}

Using the notation of Appendix \ref{appendix:likelihood}, it follows that 
\begin{align*}
e_G &\leq \left| E_\mu\{G(Z)\} - E_{\mu^p}\{G(Z)\}\right | + \left| E_{\mu^p}\{G(Z)\} - E_{\nu^{n,p}}\{G(Z_n)\} \right| \equiv B_1 + B_2
\end{align*}
and it follows from Theorem A\ref{thm:likelihood_convergence} and \eqref{eqn:hellinger_weak} that  $B_1 \leq C\psi(p)$.
Let $Z \sim \mu_0(\cdot)$ and construct the coupling $(Z,Z_n) \in V\times V_n$ through the identity $Z_n = R_n Z$.  It follows that 
\begin{align*}
B_2 &= \left | M_p^{-1} \int_V G(Z)\exp\{-\Phi^p(Z)\}\,d\mu_0 - M_{n,p}^{-1}\int_{V_n}G(Z_n) \exp\{-\Phi^p(Z_n)\}\,d\nu^{n}_0 \right | \\
&\leq M_p^{-1}\left | \int_V G(Z)\exp\{-\Phi^p(Z)\}\,d\mu_0 -  \int_{V_n}G(Z_n) \exp\{-\Phi^p(Z_n)\}\,d\nu^{n}_0 \right |\\
&\quad + \left|M_p^{-1} - M_{n,p}^{-1}\right|\int_{V_n}|G(Z_n)| \exp\{-\Phi^p(Z_n)\}\,d\nu^{n}_0 \equiv B_3 + B_4.
\end{align*}
The normalising constants $M_p$ and $M_{n,p}$ are bounded both above and below uniformly in $n$ \citep[Theorems 4.1 and 4.2,][]{stuart2010inverse}.

Let $\lambda(\cdot,\cdot)$ be the law of the coupling $(Z,Z_n)$.  Then, for any $\epsilon >0$,
\begin{align*}
M_p B_3& =\left| \int_{V\times V_n} \left[G(Z)\exp\{-\Phi^p(Z)\} - G(Z_n) \exp\{-\Phi^p(Z_n)\}\right]\,d\lambda(Z,Z_n)  \right| \\
&\leq \int_{V\times V_n} |G(Z)|\left|\exp\left\{-\Phi^p(Z)\}-  \exp\{-\Phi^p(Z_n)\right\}\right| +  \exp\{-\Phi^p(Z_n)\}\left|G(Z) - G(Z_n)\right|\,d\lambda(Z,Z_n) \\
&\leq  C \int_{V\times V_n}   \exp\left\{2C\epsilon\norm{Z}_V^2 + \epsilon\max(\norm{Z}_V^2,C\norm{Z}^2_V)\right\}\norm{Z - Z_n}_H  \,d\lambda(Z,Z_n)\\
&\leq C\left(\sup_{Z(\cdot) \in V }\frac{ \norm{Z(\cdot) - R_nZ(\cdot)}_H}{\norm{Z(\cdot)}_V} \right)\int_{V\times V_n}   \exp(3C\epsilon\norm{Z}_V^2 )\norm{Z }_V  \,d\lambda(Z,Z_n) \\
&\leq C \sup_{Z(\cdot) \in V }\frac{ \norm{Z(\cdot) - R_nZ(\cdot)}_H}{\norm{Z(\cdot)}_V},
\end{align*}
where the second inequality follows from standard bounds on the exponential function,  the assumptions on $\Phi^p(\cdot)$ and $G(\cdot)$, and the stability assumption \eqref{eqn:stability}. The third inequality follows from the observation that $Z_n= R_n Z$ almost surely, $Z(\cdot) \in V$ almost surely and  the embedding $\norm{\cdot}_H \leq C \norm{\cdot}_V$, and the final inequality follows from Fernique's theorem, which ensures that  the expectation is finite \citep{stuart2010inverse}.

To bound $B_4$, we first note that $\int_{V_n}|G(Z_n)| \exp\{-\Phi^p(Z_n)\}\,d\nu^{n}_0 < \infty$ uniformly in $n$ by assumption and Fernique's theorem.  Then it is enough to note that 
\begin{align*}
&\left|M_p^{-1} - M_{n,p}^{-1}\right| \leq \max\left(M_p^{-2}, M_{n,p}^{-2}\right)\left|M_n - M_{n,p}\right| \\
&\qquad\leq C \int_{V \times V_n}\left|\exp\{-\Phi^p(Z)\} - \exp\{-\Phi^p(Z_n)\}\right|\,d\lambda(Z,Z_n)  \leq C \sup_{Z(\cdot) \in V }\frac{ \norm{Z(\cdot) - R_nZ(\cdot)}_H}{\norm{Z(\cdot)}_V},
\end{align*}
using the reasoning above.

\end{proof}

\subsection{Convergence of the stochastic partial differential equation approximation} \label{appendix:spde_converge}

In order to apply Theorem \ref{thm:finite_dimensional} to stochastic differential equation models, it is useful to consider the abstract version of the approximation outlined in Section \ref{sec:spde}.  Let $L:H\rightarrow L^2(\Omega)$ be an operator and define the random field $Z(\cdot)$ through the equation 
$LZ(\cdot) =W(\cdot).$
Then $Z(\cdot)$ is a Gaussian random field over the sample space $H$ with covariance operator $C = L^{-1}L^{-*}$, where the star denotes the adjoint operator.   If $L_n:H\rightarrow  L^2(\Omega)$ is the Galerkin approximation to $L$ over $V_n$ defined by $
\innerprod{\phi}{L_n\psi}_H = \innerprod{\phi}{L\psi}_H$
for all $\phi,\psi \in  V_n$, then the corresponding approximate Gaussian random field $Z_n(\cdot)$ has covariance operator given by $C_n^\dag = L_n^* L_n$, where $C^\dag$ is the pseudoinverse of $C$ satisfies $C^\dag H \perp_H V_n$.  With this setup in mind, the restriction operator $R_n$ is defined by the equation $C_n = R_nCR_n^*$, from which it can be seen that $R_n = L_n^\dag L$ is a natural choice. If $Z_n(\cdot)$ converges in distribution to $Z(\cdot)$, which is the case for the models in Section \ref{sec:spde} \citep{Lindgren2011}, we can use Skorohod's representation theorem to construct, possibly on a different probability space, the coupling $(Z,Z_n)$, defined by $Z_n(\cdot) = R_nZ(\cdot)$ almost surely, that is required in Theorem A\ref{thm:finite_dimensional}. Hence 
\begin{equation} \label{eqn:pde_error}
\sup_{Z(\cdot) \in V}\frac{ \norm{Z(\cdot) - R_nZ(\cdot)}_H}{\norm{Z(\cdot)}_V} =\sup_{f(\cdot) \in LV}\frac{ \norm{L^{-1}f(\cdot) - L_n^{\dag}g(\cdot)}_H}{\norm{L^{-1}f(\cdot)}_V}
\end{equation}
and the rate of convergence is governed by how well solutions to the partial differential equation $Lx(\cdot) = f(\cdot)$ can be approximated by solutions to $L_n x_n(\cdot) = f(\cdot)$.

The following Theorem shows that, for fixed parameters, the approximate posterior distributions computed using the stochastic partial differential equation approach introduced by \citet{Lindgren2011} converge. 

\begin{theorem} \label{thm:spde_converge}
Let $\Omega \in \mathbb{R}^2$ be a convex polygon. Let $G$ be a Lipschitz continuous function that satisfies the assumptions of Theorem A\ref{thm:finite_dimensional}. Assume that $\kappa >0$ and the family of triangulations $\mathcal{T}_n$ is quasi-uniform \citep[Definition 4.4.13][]{book107}.
Then, if the approximate posterior measure $\nu^{n,p}$  is defined using the approximation and the integration rule outlined in Section \ref{sec:spde}, then, for any $\epsilon >0$, $
e^p_G = \left| E_{\mu^p}\{G(Z)\} - E_{\nu^{n,p}}\{G(Z_n)\} \right| \leq Ch^{1-\epsilon},
$
where $h$ is the length of the largest edge in the mesh. 
\end{theorem}
\begin{proof}

The use of Theorem A\ref{thm:finite_dimensional} is  complicated by the lack of Sobolev regularity of the Gaussian random field.  In particular, the field $Z(s)$ considered in Section \ref{sec:spde} is almost surely in $V = H^{1-\epsilon}(\Omega)$ for all $\epsilon>0$ \citep[Lemma 6.2.7][]{stuart2010inverse}.  We then take $V = L^2(\Omega)$ and define the differential operator as $L = \kappa^2 - \Delta$.   We define the approximation space $V_n$ to be the space of piecewise linear functions defined over the triangulation $T_n$ and let $h$ be the maximum edge length.  Under the assumptions on $\Omega$, $LV = H^{-1-\epsilon}(\Omega)$ \citep[3.12]{ErnGuermond}, where a Sobolev space with negative index is defined as the dual of the space with the corresponding positive index. This is consistent with the fact that white noise can be considered a random function in $H^{-1-\epsilon}(\Omega)$ \citep{walsh1986introduction}.   In order to define $L_n$, we need the $L^2(\Omega)$--orthogonal projector $P_n:H\rightarrow V_n\equiv \mathbb{R}^n$ and we  define the Galerkin approximation as $L_n^{-1} = P_n^* (\kappa^2 C_n + G_n - B_n)^{-1}P_n$.   

Fix $\epsilon \in(0,\frac{1}{2})$ and $f \in H^{-1-\epsilon}$ and let $z$ be the distributional solution to $Lz = f$.  We emphasise that $f(\cdot)$ is not a function in an ordinary sense, but 
 a distribution, and in the remainder of this proof integrals containing $f \in H^{-s}(\Omega)$ ($s>0$)  should be interpreted as $
\int_\Omega f(s)\phi(s)\,ds \equiv \innerprod{f}{\phi}_{H^{-s}(\Omega),H^s(\Omega)},
$ where the angle brackets denote the duality pairing.

As the standard convergence theory for finite element methods \citep{book107,ErnGuermond}, would require the sample paths to be almost surely in $H^2(\Omega)$, we modify the arguments used to prove Proposition 1 in   \citet{scott1976optimal}.    The crucial step in Scott's method is to approximate $f(s)$ by a  piecewise linear function $f_n(s)$ defined over $T_n$ such that $\norm{f_n}_{L^2(\Omega)}$ is controlled by a negative power of $h$.  We define $f_n(s)$ as $
\int_\Omega f_n(s) v_n(s) \,ds = \int_\Omega f(s)v_n(s)\,ds$
for all $v_n \in V_n$, 
where the second integral is understood in the sense of distributions and makes sense because $V_n \subset H^{1+\epsilon}(\Omega)$ \citep{belgacem2001some}.  For an arbitrary $v \in L^2$, let $v_n \in V_n$ be the orthogonal projection of $v$ onto $V_n$.  Then 
$$
\int_\Omega f_n(s) v(s) \,ds = \int_\Omega f(s)v_n(s)\,ds \leq \norm{f}_{H^{-1-\epsilon}(\Omega)}\norm{v_n}_{H^{1+\epsilon}(\Omega)} 
\leq Ch^{-1-\epsilon} \norm{f}_{H^{-1-\epsilon}(\Omega)}\norm{v}_{L^2(\Omega)},$$
where the final inequality follows from equations (1.5) and (1.6) of \citet{belgacem2001some}.  As $v$ was arbitrary, this gives an appropriate bound for $\norm{f_n}_{L^2(\Omega)}$.  

Define $\tilde{z}(s) \in H^2(\Omega)$ to be the solution of $
\int_\Omega \nabla \tilde{z}(s) \nabla \phi(s) \,ds = \int_\Omega f_n(s)\phi(s)\,ds$
for all $\phi \in H^1(\Omega)$ and consider the finite element approximation $z_n \in V_n$ defined as  $
\int_\Omega \nabla {z}_n(s) \nabla \phi_n(s) \,ds = \int_\Omega f_n(s)\phi_n(s)\,ds$
for all $\phi_n \in V_n.$
The dependence of $\tilde{z}$ on $n$ is suppressed for  readability. The key observation is that $z_n(s)$ can be considered a finite element approximation to both $z(s)$ and $\tilde{z}(s)$ as $ \int_\Omega f_n(s)\phi_n(s)\,ds =  \int_\Omega f(s)\phi_n(s)\,ds$ for every $\phi_n\in V_n$.  It follows from standard finite element theory \citep[Theorem 5.7.6]{book107}  that
$$
\norm{\tilde{z} - z_n}_{L^2(\Omega)} \leq Ch^2 \norm{\tilde{z}}_{H^2(\Omega)}
\leq Ch^2 \norm{f_n}_{L^2}(\Omega) 
\leq Ch^{1-\epsilon}.
$$

The final ingredient of the proof is to bound $\norm{z - \tilde{z}}_{L^2(\Omega)}$.  Fix $\phi(s) \in L^2(\Omega)$ and let $\Phi(s)$ be the solution to $L^*\Phi = \phi$, where $L^*$ is the adjoint of $L$.  Then it follows that, for any $v_n \in V_n$,
\begin{align*}
\int_\Omega \{z(s) - \tilde{z}(s)\}\phi(s)\,ds &= \int_\Omega  \{z(s) - \tilde{z}(s)\}L^*\Phi(s)\,ds =   \int_\Omega  L\{z(s) - \tilde{z}(s)\}\Phi(s)\,ds \\
&= \int_\Omega  \{f(s) - f_n(s)\}(\Phi(s)- v_n )\,ds \\
&\leq C\norm{f}_{H^{-1-\epsilon}(\Omega)} h^{-1-\epsilon}\inf_{v_n \in V_n}\left(  h^{1+\epsilon}\norm{\Phi - v_n}_{H^{1+\epsilon}(\Omega)} +\norm{\Phi-v_n}_{L^2(\Omega)} \right)\\
&\leq C\norm{f}_{H^{-1-\epsilon}(\Omega)} h^{-1-\epsilon}h\norm{\Phi}_{H^2(\Omega)} \leq C\norm{f}_{H^{-1-\epsilon}(\Omega)} h^{1-\epsilon}\norm{\phi}_{L^2(\Omega)},
\end{align*}
where the second line follows from the orthogonality of $f-f_n$ to $V_n$, the penultimate inequality follows from Theorem 14.4.2 of \citet{book107} and the fact that $\Phi \in H^2(\Omega)$.  As $\phi(s)$ was arbitrary, this completes the proof.
\end{proof}
Unfortunately we are unable to prove that the entire posterior distribution converges.  This is due to the gap in the theory identified in Remark A\ref{remark:gap}, which prevents Theorem A\ref{thm:likelihood_convergence} from giving the rate $h^{-p}$ where $Z(s) \in H^{1-\epsilon}(\Omega)$.  However, if it is true that $E_{\mu_0} [\norm{\exp\{Z(\cdot)\}}_{H^\gamma(\Omega)}]$ is bounded, then the observation that the integration scheme considered in Section \ref{sec:spde} has $\mathcal{O}(h)$ error leads to the following conjecture.
\begin{conjecture}
Under the conditions of Theorem A\ref{thm:spde_converge},  for any $\epsilon >0$, $$
e_G = \left| E_{\mu}\{G(Z)\} - E_{\nu^{n,p}}\{G(Z_n)\} \right| \leq Ch^{1-\epsilon}.
$$
\end{conjecture}

\subsection{Higher order schemes} \label{appendix:higher_order}

In this section, we sketch a method that provides higher order convergence whenever the Gaussian random field  is sufficiently smooth. We use the truncated Karhunen--Lo\`{e}ve expansion as our finite-dimensional approximations to $Z(\cdot)$.  Let $\Omega = [ -\pi,\pi]^d$ and construct a lattice over $\Omega$ with $N$ partitions in each dimension.  If $f(s) \in H^\gamma(\Omega)$, there is a tensor product Gaussian quadrature rule with $\left\lceil\{(\gamma-1)/2\}\right\rceil^d$ points in each lattice cell such that  $
\left| \int_{\Omega} f(s)\,ds - \sum_{i=1}^p \tilde{\alpha}_i f(\tilde{s}_i) \right| \leq C N^{-\gamma} \norm{f(\cdot)}_{H^\gamma(\Omega)}.
$  Let $
Z_\alpha = \sum_{j\in\mathbb{N}^d} \lambda_j^{\alpha/2} z_j \psi_j(s)
$, where $\{(\lambda_j,\psi_j(s))\}_{j\in\mathbb{N}^d}$ are the eigenvalues and eigenfunctions of $\kappa^2 - \Delta$ on the domain $\Omega$, $z_j \stackrel{\text{i.i.d.}}{\sim} N(0,1)$, and $\mathbb{N}$ is the set of non-negative integers.  $Z_\alpha(\cdot) \in H^{\alpha-d/2}$ almost surely \citep[Lemma 6.27]{stuart2010inverse}.   Let $Z_\alpha^N(\cdot)$ be the Gaussian random field where the Karhunen--Lo\`{e}ve expansion is now summed over $[0,N]^d$.  

\begin{corollary}
Assume that $\alpha-d/2$ is an integer and let $\nu^{n,p}(\cdot)$ be the approximate posterior distribution computed using the integration scheme and the truncated Karhunen--Lo\`{e}ve expansion $Z_\alpha^N(\cdot)$ and let $\mu(\cdot)$ be the true posterior distribution computed using the exact log-Gaussian Cox process likelihood and the infinite-dimensional Gaussian random field $Z_\alpha(\cdot)$.  Then, under the conditions on $G$ outlined in Theorem A\ref{thm:finite_dimensional}, $
e_G = | E_\mu\{G(Z)\} - E_{\nu^{n,p}} \{G(Z_n)\}| \leq CN^{-t}
$ for every $t < \alpha -d/2$.
\end{corollary}
\begin{proof}
The proof follows  from Theorems A\ref{thm:likelihood_convergence}, A\ref{thm:finite_dimensional}, and Theorem 4.2 of \citet{dashti2011uncertainty}.
\end{proof}

\end{document}